\documentclass[aps,twocolumn,reprint,showpacs]{revtex4-1} %
\pdfoutput=1
\usepackage{graphics,graphicx,amsfonts,amsmath,amsbsy,amssymb,color}
\usepackage[singlelinecheck=false]{subfig}
\usepackage{verbatim}  
\usepackage{psfrag}  
\usepackage{tabularx}
\usepackage{xmpmulti}
\usepackage[justification=justified]{caption}
\usepackage{hyperref}

\newcommand{\reffig}[1]{{Fig.~\ref{#1}}}
\newcommand{\reftab}[1]{{Table~\ref{#1}}}
\newcommand{\refsec}[1]{{Sec.~\ref{#1}}}
\newcommand{\appropto}{\mathrel{\vcenter{
  \offinterlineskip\halign{\hfil$##$\cr
    \propto\cr\noalign{\kern2pt}\sim\cr\noalign{\kern-2pt}}}}}

\hypersetup{hyperindex,breaklinks,colorlinks=true,linkcolor=blue,citecolor=blue,urlcolor=magenta,filecolor=blue}

\begin{document}

\title{Coupled Cluster Channels in the Homogeneous Electron Gas}

\author{James~J.~Shepherd}
\email{jjs6@rice.edu}
\affiliation{Department of Chemistry and Department of Physics and Astronomy, Rice University, Houston, TX 77005-1892}
\author{Thomas~M.~Henderson}
\affiliation{Department of Chemistry and Department of Physics and Astronomy, Rice University, Houston, TX 77005-1892}
\author{Gustavo~E.~Scuseria}
\affiliation{Department of Chemistry and Department of Physics and Astronomy, Rice University, Houston, TX 77005-1892}

\begin{abstract}
We discuss diagrammatic modifications to the coupled cluster doubles (CCD) equations, wherein different groups of terms out of rings, ladders, crossed-rings and mosaics can be removed to form approximations to the coupled cluster method, of interest due to their similarity with various types of random phase approximations. The finite uniform electron gas is benchmarked for 14- and 54-electron systems at the complete basis set limit over a wide density range and performance of different flavours of CCD are determined. These results confirm that rings generally overcorrelate and ladders generally undercorrelate; mosaics-only CCD yields a result surprisingly close to CCD. We use a recently developed numerical analysis [\href {http://dx.doi.org/10.1103/PhysRevLett.110.226401} {J. J. Shepherd and A. Gr\"uneis, Phys. Rev. Lett. {\bf 110}, 226401 (2013) }] to study the behaviours of these methods in the thermodynamic limit. We determine that the mosaics, on forming the Brueckner one-body Hamltonian, open a gap in the effective one-particle eigenvalues at the Fermi energy. Numerical evidence is presented which shows that methods based on this renormalisation have convergent energies in the thermodynamic limit including mosaic-only CCD, which is just a renormalised MP2. All other methods including only a single channel, namely ladder-only CCD, ring-only CCD and crossed-ring-only CCD, appear to yield divergent energies; incorporation of mosaic terms prevents this from happening.
\end{abstract}
\date{\today}

\maketitle

\section{Introduction}

It has recently been discussed by Scuseria \emph{et al.}~\cite{scuseria_ground_2008,scuseria_particle-particle_2013} and Yang and coworkers~\cite{peng_equivalence_2013,van_aggelen_exchange-correlation_2013} that the random phase approximation and coupled cluster doubles bear a formal connection to one another, linking two widely popular methods. 
This relates to the well-known observation that the amplitude equations can be separated into sets of terms grouped together by what they represent in diagrammatic many-body perturbation theory~\cite{cizek_correlation_1966}. Rings represent particle-hole contractions, ladders represent particle-particle and hole-hole contractions, what we term mosaics involve joint ladder and ring contractions and renormalise the eigenvalues of the one-particle Hamiltonian, and crossed-rings are the exchange components of the Coulomb ring diagrams needed to maintain the antisymmetry of the amplitudes. These offer the possibility of a modified electronic structure approximation that takes advantage of intuitions concerning these diagrams. 

The aim of this paper is to explore these developments further through their application to the homogeneous electron gas model in three dimensions. This is the simplest fully periodic condensed matter system and has been the subject of intense investigation over many years. In recent times, the study of the gas with modern, numerical quantum chemical methods employing finite electron numbers and basis sets has been made substantially easier. 
Full configuration interaction quality benchmarks for finite basis sets have become available~\cite{shepherd_full_2012,shepherd_investigation_2012} and the relationship between these and the complete basis set limit for plane waves has been thoroughly analysed~\cite{shepherd_full_2012,shepherd_convergence_2012,shepherd_many-body_2013,gruneis_explicitly_2013,marsman_second-order_2009}. This understanding forms a bridge between finite basis set work and the energies obtained from diffusion Monte Carlo, whose complete-basis-set ground-state energies have provided a wide range of very accurate benchmark data~\cite{lopez_rios_inhomogeneous_2006,kwon_effects_1998,holzmann_backflow_2003,ortiz_correlation_1994,ceperley_ground_1980}. In spite of the wealth of historical literature on the electron gas, there are still new and interesting recent studies on, for instance, the effects of spin-polarisation on correlation factors~\cite{spink_quantum_2013}, determination of Landau Fermi liquid theory parameters~\cite{drummond_quantum_2013,drummond_diffusion_2013} and finite-temperature simulations on the warm-dense gas~\cite{brown_exchange-correlation_2013}.

Using judicious subsets of diagrams is a common theme in many-body perturbation theory. This is especially true for the three dimensional electron gas, where an inappropriate choice of diagrams can easily lead to a divergence in the energy due to problematic terms appearing at every finite order of perturbation theory~\cite{macke_uber_1950,gell-mann_correlation_1957,mattuck_guide_1992,hirata_thermodynamic_2012}. This was circumvented by Gell-Mann and Brueckner, who calculated the correlation energy for the high-density limit of the electron gas using an infinite resummation of ring diagrams~\cite{gell-mann_correlation_1957}. This corresponded to the so-called random phase approximation (RPA) energy for the gas described some years earlier by Bohm and Pines~\cite{bohm_collective_1953,pines_collective_1952,bohm_collective_1951}. Somewhat later, Freeman added to this the idea of finite-order screened exchange~\cite{freeman_coupled-cluster_1977}. Bishop and L\"uhrman, at roughly the same time, extensively examined the effect of adding various diagrams into equations built out of a coupled cluster ansatz looking more like modern RPA with linearised ladder terms~\cite{bishop_electron_1978,bishop_electron_1982,bishop_overview_1991}. These ideas seem to have been investigated intermittently since then~\cite{bishop_pairing_1988,bishop_pairing_1988-1}. 

In the intervening years, the random phase approximation has become a popular and routine electronic structure method~\cite{furche_molecular_2001,heselmann_random-phase_2011,furche_developing_2008,furche_developing_2008-1,eshuis_fast_2010,eshuis_basis_2012,paier_hybrid_2010,janesko_long-range-corrected_2009,janesko_long-range-corrected_2009-1,henderson_connection_2010,irelan_long-range-corrected_2011,janesko_role_2009}; for more details the reader is directed towards a selection of reviews~\cite{eshuis_electron_2012, ren_random-phase_2012,paier_assessment_2012}. Some of these focus on the idea that rings describe long-range correlation well~\cite{dobson_asymptotics_2006,harl_accurate_2009,lebegue_cohesive_2010,toulouse_closed-shell_2011,angyan_correlation_2011,toulouse_adiabatic-connection_2009,irelan_long-range-corrected_2011,janesko_long-range-corrected_2009}. In parallel with this there has been long-standing discussion of the use of ladders for short-range correlation and how it describes the pair correlation function~\cite{calmels_pair-correlation_1998,nagano_correlations_1984,yasuhara_short-range_1972,yasuhara_electron_1974,yasuhara_erratum:_1974,lowy_electron_1975,bedell_short-range_1978,qian_-top_2006,drummond_quantum_2009,awa_electron_1980,yasuhara_new_1988}. It is also interesting to remark that developments are still being explored in the nuclear physics community~\cite{hagen_coupled-cluster_2013}.

There has also been a recent rise in interest in treating solid state problems with quantum chemical wavefunction theories~\cite{paier_accurate_2009,dovesi_hartreefock_1984,gillan_high-precision_2008,pisani_periodic_2008,marsman_second-order_2009,schwerdtfeger_convergence_2010,maschio_fast_2007,usvyat_approaching_2011,hirata_coupled-cluster_2004,gruneis_second-order_2010,pino_importance_2004,del_ben_second-order_2012,geckeis_mineralwater_2013,del_ben_electron_2013,usvyat_linear-scaling_2013,muller_incrementally_2013}, and in particular coupled cluster~\cite{schwerdtfeger_convergence_2010,muller_wavefunction-based_2012,nolan_comparison_2010,gruneis_natural_2011,booth_towards_2013,shepherd_many-body_2013,shepherd_convergence_2012,gruneis_explicitly_2013,keceli_fast_2010,hirata_coupled-cluster_2004}, where authors are attempting to build on its success for molecular systems~\cite{bartlett_coupled-cluster_2007}. 

We aim here to more extensively explore the connection between coupled cluster on the electron gas and diagrammatic ideas. We will predominantly use ground-state energy estimates as our method of discussing these theories. The rest of this paper is organised as follows. We start by introducing our model system and discussing technical considerations, then benchmark the 14 electron problem at the complete basis set limit with the different `flavours' of CCD. We show that methods involving ring diagrams generally overcorrelate in the electron gas and methods involving ladder diagrams generally undercorrelate. In other words, the electron gas behaves in common with previous observations in different systems~\cite{bartlett_many-body_1974,kelly_many-body_1966,kelly_many-body_1964,kelly_correlation_1963,ostlund_perturbation_1975}. We then examine the applicability of these findings to the thermodynamic limit and discuss the role of mosaics in producing an insulating reference state and circumventing divergences. Throughout, we will discuss the extent to which the conclusions we draw have been already present in the literature, and our aim is to produce a modern perspective on these views. This paper sets the scene to a separate paper where we discuss range separating CCD to try to achieve higher accuracy energies for the electron gas~\cite{shepherd_range_2013}.

\section{The electron gas}

The Hamiltonian for the $N$-electron, or simulation-cell, homogeneous electron gas reads
\begin{equation}
\hat{H}=T+V_\textrm{ee}+V_\textrm{eb}+V_\textrm{bb}%
\end{equation}
where these terms are the kinetic energy operator for the electrons, the electron-electron interaction, the electron-background interaction and the background-background interaction.
As is well known, there are difficulties working within periodic systems for Coulomb interactions. In the electron gas, many terms cancel out and we can take advantage of these cancellations to express the four-index integrals in a plane wave basis set as
\begin{align}
v_{ab}^{ij}&=\langle i j  | a b \rangle \\
&= v ( {\bf k}_a -{\bf k}_i  ) \delta_{{\bf k}_a -{\bf k}_i,{\bf k}_j -{\bf k}_b}, \label{matrixel}\\
v({\bf q}) &= \left\{
\begin{array}{ll}
\frac{1}{L^3}  \frac{4\pi}{{\bf q}^2}, & {\bf q}\neq\bf{0} \\
v_M, & \mbox{{\bf q}=\bf{0}}.
\end{array}
\right. %
\end{align}
and the Hartree--Fock eigenvalues are
\begin{align}
\epsilon_i &= \frac{1}{2} {\bf k}_i^2 - \sum_{j \in \textrm{occ}} \langle i j |   j i \rangle \\
&= \left\{
\begin{array}{ll}
\frac{1}{2} {\bf k}_i^2 - \sum_{\substack{j \in \textrm{occ}\\i \neq j}}  v ( {\bf k}_i -{\bf k}_j  ) - v_M , &\quad i \in \textrm{occ} \\
\frac{1}{2} {\bf k}_i^2 - \sum_{\substack{j \in \textrm{occ}}}  v ( {\bf k}_i -{\bf k}_j  ) , &\quad i \in \textrm{virt}.
\end{array}
\right. %
\end{align}
Here, a finite box size of length $L$ quantises ${\bf q}= \frac{2 \pi }{L} {\bf n}$ and the zero momentum integral (${\bf k}_a ={\bf k}_i$) is given by $v_M$, the Madelung constant, which is a feature of the finite electron gas which vanishes in the thermodynamic limit found as $N\rightarrow\infty$ and $L\rightarrow\infty$. The Madelung constant is defined uniquely for a cell geometry and box length, and here we calculate it to be $v_M L\approx 2.8372$ where $L$ is the length of the box. 
The box length is uniquely defined by the electron number $N$ and the density parameter $r_s$, the radius that on average encloses one electron. Another parameter required for finite basis methods such as coupled cluster is a way of defining the basis set used. This is typically a kinetic energy cutoff, and we will use the number of plane-wave spin orbitals, $M$, to express the size of our basis set. Recently, quantum Monte Carlo benchmarks have become available for the finite basis sets~\cite{shepherd_full_2012,shepherd_investigation_2012}, which are useful for benchmarking quantum chemical methods~\cite{shepherd_convergence_2012,gruneis_explicitly_2013,shepherd_many-body_2013,roggero_quantum_2013}, and a simple extrapolation scheme to obtain complete basis set limit energies is well-understood~\cite{shepherd_convergence_2012}. Some authors are attempting to use coupled cluster wave functions to guide quantum Monte Carlo simulations~\cite{roggero_quantum_2013}.

The simplicity of this model is one of the reasons that it has been extremely well studied. The plane-wave basis is the canonical RHF basis and all one and two-electron integrals have analytic forms, so they need not be stored. Momentum symmetry means amplitude storage is also reduced by a factor of $M$ compared with other coupled cluster codes, and the computational cost scales as $\mathcal{O} \left( M^4 \right)$. Furthermore, there is a rigorous absence of single excitations in the plane-wave HEG, caused by momentum symmetry; single excitations are disallowed from the exact wavefunction because they necessarily have a different total momentum quantum number than the reference state. This means that coupled cluster singles and doubles (CCSD) and coupled cluster doubles (CCD) are equivalent. The plane-wave basis set diagonalises the Fock matrix, the one-body density matrix and also the one-body Brueckner Hamiltonian~\cite{bethe_nuclear_1956,brueckner_two-body_1954,brueckner_nuclear_1954}. The link between the Brueckner Hamiltonian and the mosaic terms will be discussed explicitly at the end of \refsec{IIImodCCD}.

Our aim is to produce complete basis set coupled cluster energies and compare them with the most accurate data available from quantum Monte Carlo simulations. The technical details of our calculations are as follows. Coupled cluster calculations are performed on finite basis sets of size $M$ and then extrapolated to the complete basis set (CBS) limit using a $1/M$ direct extrapolation as described in Ref.~\onlinecite{shepherd_investigation_2012} and used in a variety of studies~\cite{gruneis_making_2009,gruneis_second-order_2010,harl_cohesive_2008,harl_assessing_2010,marsman_second-order_2009,shepherd_investigation_2012,gruneis_second-order_2010,shepherd_full_2012,del_ben_electron_2013}. We note that it is also now possible to use F12 corrections in plane wave basis sets~\cite{gruneis_explicitly_2013}, but do not use these here. The convergence of the modified CCD equations is aided by the use of DIIS~\cite{scuseria_accelerating_1986}. We study in particular $N=14$ for $0.1 \leq r_s \leq 100.0$ and $N=54$ for $0.5 \leq r_s \leq 20.0$, for which there is high-quality benchmark data available. We draw benchmark data from a number of quantum Monte Carlo (QMC) sources. For $N=14$ at the high and metallic densities, $0.5 \leq r_s \leq 5.0$, we make use of CBS data from full configuration interaction QMC~\cite{booth_fermion_2009,cleland_communications:_2010,booth_breaking_2011,shepherd_full_2012,booth_towards_2013}, which have been already published~\cite{shepherd_investigation_2012}. We supplement these with diffusion Monte Carlo results at $r_s=10.0, 20.0, 50.0$ and 100.0~a.u. obtained from collaborators~\cite{PabloPersComm,needs_continuum_2010}, which will be published elsewhere~\cite{shepherd__2013}. The data for $N=54$ come from a previous diffusion Monte Carlo study~\cite{lopez_rios_inhomogeneous_2006}. These have been combined into a single QMC data set that are used in our plots below, and represent the highest-accuracy results to date for these systems. These benchmarks are for the purposes of this study exact. Full configuration interaction quantum Monte Carlo suffers from an initiator error~\cite{cleland_communications:_2010,booth_breaking_2011}, which can be very effectively reduced or removed for the electron gas~\cite{shepherd_investigation_2012}. Diffusion Monte Carlo suffers from a fixed-node error, which reduces at lower densities where it becomes relatively small~\cite{foulkes_quantum_2001,kwon_effects_1998,ceperley_ground_1980}.
  
\section{Modified CCD equations}
\label{IIImodCCD}

The central premise of this study lies in modification of the amplitude equations from CCD to form new CCD-like approximations. For a canonical basis which diagonalises the Fock matrix, the CCD amplitude equations read
\begin{equation}
\begin{split}
(\epsilon_i +\epsilon_j &-\epsilon_a -\epsilon_b) t_{ij}^{ab}=\bar{v}_{ij}^{ab} \\
&+\frac{1}{2} \bar{v}_{cd}^{ab} t_{ij}^{cd} +\frac{1}{2} \bar{v}_{ij}^{kl}  t_{kl}^{ab} + \frac{1}{4} \bar{v}_{cd}^{kl} t_{ij}^{cd} t_{kl}^{ab} \\
&+\bar{v}_{cj}^{kb} t_{ik}^{ac} + \bar{v}_{ci}^{ka} t_{jk}^{bc} + \bar{v}_{cd}^{kl} t_{lj}^{db} t_{ik}^{ac} \\
&- \bar{v}_{cj}^{ka} t_{ik}^{bc} - \bar{v}_{ci}^{kb} t_{jk}^{ac} - \bar{v}_{cd}^{kl} t_{lj}^{da} t_{ik}^{bc} \\
&+\frac{1}{2} \bar{v}_{cd}^{kl} \left[  t_{lj}^{ab} t_{ik}^{cd} - t_{li}^{ab} t_{jk}^{cd} + t_{ji}^{db} t_{kl}^{ac} -t_{ij}^{da} t_{kl}^{bc} \right]
\end{split}
\end{equation}
where in these equations, $\bar{v}_{ab}^{ij}={v}_{ab}^{ij}-{v}_{ba}^{ij}=\langle ij || ab \rangle$, $\epsilon$ are the Hartree--Fock eigenvalues and repeated indices are summed on the right-hand side. Hole states are labelled with $i, j, k, $ and $l$  and particle states are labelled with $a, b, c$, and $d$. This expression is grouped such that the top line represents the driver term, the second line ladder terms (l), the third line ring (r) terms, the fourth line crossed-ring (x) terms and the final line mosaics (m). 

For simplicity, it is possible to represent the amplitude equations schematically as follows:
\begin{equation}
\begin{split}
0=&\,\textrm{driver}+\textrm{rings}+\textrm{crossed-rings}\\&+\textrm{ladders}+\textrm{mosaics}.
\label{schematic}
\end{split}
\end{equation}

Our idea is to allow each channel to be independently added into or excluded from a CCD calculation. 
This creates a new range of methods and potentially extends the flexibility of CCD. 
These choices are inspired by the connections to the RPA discussed in Refs. \onlinecite{scuseria_ground_2008} and \onlinecite{scuseria_particle-particle_2013}.
In this paper the different channels will be prefixed to the abbreviation CCD (for example rCCD denotes CCD with just the rings channel and driver term). We note in passing that the modification of the CCD equations to approximately capture higher-order correlations has been discussed recently in various contexts~\cite{neese_efficient_2009,meyer_ionization_1971,paldus_approximate_1984,bartlett_addition_2006,huntington_pccsd:_2010,masur_efficient_2013}. In particular, we note that Kats and Manby have attempted to remove a quadratic term due to ring and crossed-ring diagrams~\cite{kats_communication:_2013}, and follow a parameterisation by Huntingdon and Nooijen to restore the accuracy of their resultant equations for two-electron systems~\cite{huntington_pccsd:_2010}. From this is it also well-known that the smallest set of diagrams required to retain accuracy for two-electron systems using Brueckner orbitals is the hole-hole component of the mosaic terms and quadratic ladder terms~\cite{huntington_pccsd:_2010}. 

Each combination of the channels results in different (approximate) amplitudes for CCD which may yield a higher or lower non-variational energy
\begin{equation}
E_\text{corr}=\frac{1}{4} t_{ij}^{ab} \bar{v}_{ab}^{ij}.
\label{ecorrCCD}
\end{equation}

Mosaic terms are slightly different to the rest of the diagrams because they serve to renormalise the eigenvalues of the Brueckner one-body Hamiltonian~\cite{scuseria_connections_1995}:
\begin{align}
\tilde{F}_{il} &= \epsilon_i \delta_{il} - \frac{1}{2} \bar{v}_{cd}^{kl} t_{i k}^{cd}  \label{B1H1} \\
\tilde{F}_{da} &= \epsilon_a \delta_{ad } + \frac{1}{2} \bar{v}_{cd}^{kl}  t_{kl}^{ac}  \label{B1H2}
\end{align}
Here, repeated indices only denote summation in terms not involving $\epsilon$.
 It is important to note that here we use the definition of the Brueckner one-body Hamiltonian discussed in Refs.~\onlinecite{scuseria_connections_1995} and \onlinecite{bartlett_towards_2009}.
Using just the driver term and the mosaics, \emph{i.e.} mCCD, reduces to a kind of ``self-consistent'' Brueckner MP2. 

 To be clear about what we mean by this self-consistent MP2, note that the mosaic-only CCD amplitude equations can be cast as
\begin{equation}
F_i^k t_{kj}^{ab} + F_j^k t_{ik}^{ab} - F_c^a t_{ij}^{cb} - F_c^b t_{ij}^{ac} = \bar{v}_{ij}^{ab}.
\end{equation}
This is the same as the amplitude equations in MP2 except that the effective Hamiltonian elements $F_p^q$ are given by Eqs. \ref{B1H1} and \ref{B1H2}
rather than the usual Hartree-Fock values; because these elements depend on the $t$ amplitudes which, in turn, depend on these elements, the whole looks in practice like a self-consistent MP2 even though the method is fundamentally infinite order.

This is particularly clear for the case of the HEG, where $\mathbf{F}$ is diagonal in the plane wave basis and we have simply
\begin{align}
t_{ij}^{ab} &= \frac{\bar{v}_{ij}^{ab}}{\epsilon_i^\mathrm{B} + \epsilon_j^\mathrm{B} - \epsilon_a^\mathrm{B} - \epsilon_b^\mathrm{B}},
\\
\epsilon_i^\mathrm{B} &= \epsilon_i^\mathrm{HF} + \frac{1}{2} \sum_{lcd} \bar{v}^{il}_{cd} t_{il}^{cd},
\\
\epsilon_a^\mathrm{B} &= \epsilon_a^\mathrm{HF} - \frac{1}{2} \sum_{kld} \bar{v}^{kl}_{ad} t_{kl}^{ad}.
\end{align}
Here we have made clear with superscripts the distinction between Hartree--Fock eigenvalues and Brueckner eigenvalues.
Where a plane wave with momentum $k$ has energy $\frac{1}{2} k^2$, Hartree-Fock dresses the kinetic energy to account for the effects of exchange, yielding $\epsilon^\mathrm{HF}$.  One can think of this mosaic-only CCD as also incorporating correlation effects in assigning a single-particle energy to a plane wave.
 
 The Brueckner Hamiltonian has been discussed in the context of Brueckner coupled cluster~\cite{watts_coupled-cluster_1994,sherrill_energies_1998,hampel_comparison_1992,scuseria_connections_1995,scuseria_alternative_1994} and Brueckner RPA~\cite{moussa_cubic-scaling_2014}.

\section{Application to 14-electron HEG}

We now benchmark the 14 electron HEG with a variety of CCD-like methods. In this section, we focus on a presentation of our own numerical data, having summarised similar work by other authors in the introduction: more comparison is made in \refsec{V}. We begin with a discussion of the rings, crossed-rings and RPA. The rings-only amplitude equations read
\begin{equation}
\begin{split}
0=\, &\bar{v}_{ij}^{ab}+t_{ij}^{ab}(\epsilon_a+\epsilon_b-\epsilon_i-\epsilon_j) \\
&+\bar{v}_{cj}^{kb} t_{ik}^{ac} + \bar{v}_{ci}^{ka} t_{jk}^{bc} + \bar{v}_{cd}^{kl} t_{lj}^{db} t_{ik}^{ac} .
\label{eq:rings}
\end{split}
\end{equation}
Rings-only CCD is equivalent to particle-hole RPA when the above equations are solved, except that a different energy expression is used~\cite{scuseria_ground_2008},
\begin{equation}
E_\text{corr}=\frac{1}{2} t_{ij}^{ab} \bar{v}_{ab}^{ij},
\end{equation}
and this yields what is commonly called the dRPA+SOSEX energy. Removal of the SOSEX term, which involves removing the anti-symmetrisation from the expression above, yields the dRPA energy. The factor of a half comes from the plasmon formula for the dRPA energy.

\begin{figure*}
    \subfloat[\label{fig1}]{%
      \includegraphics[width=0.46\textwidth]{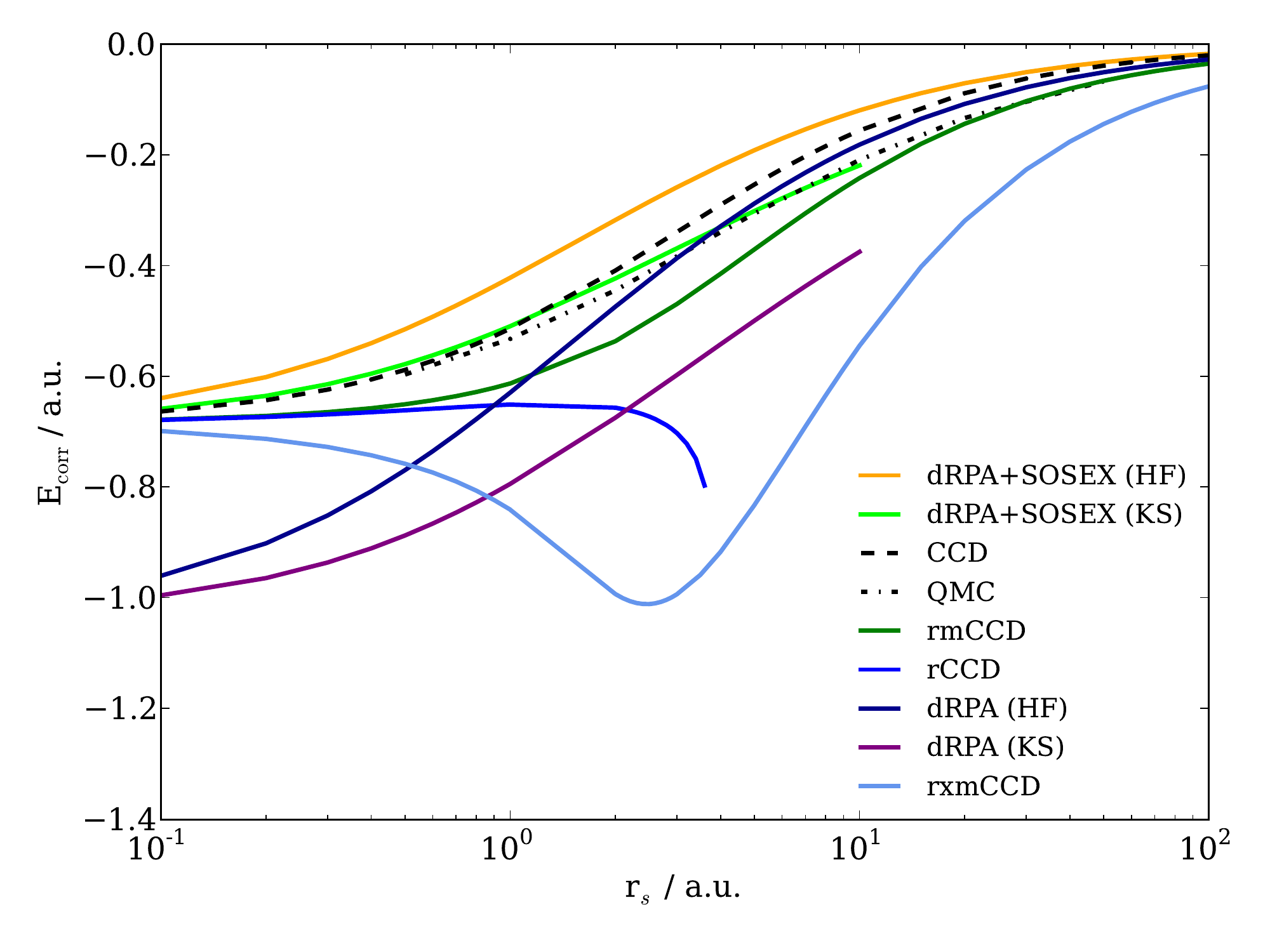}
    }
    \subfloat[\label{fig2}]{%
      \includegraphics[width=0.46\textwidth]{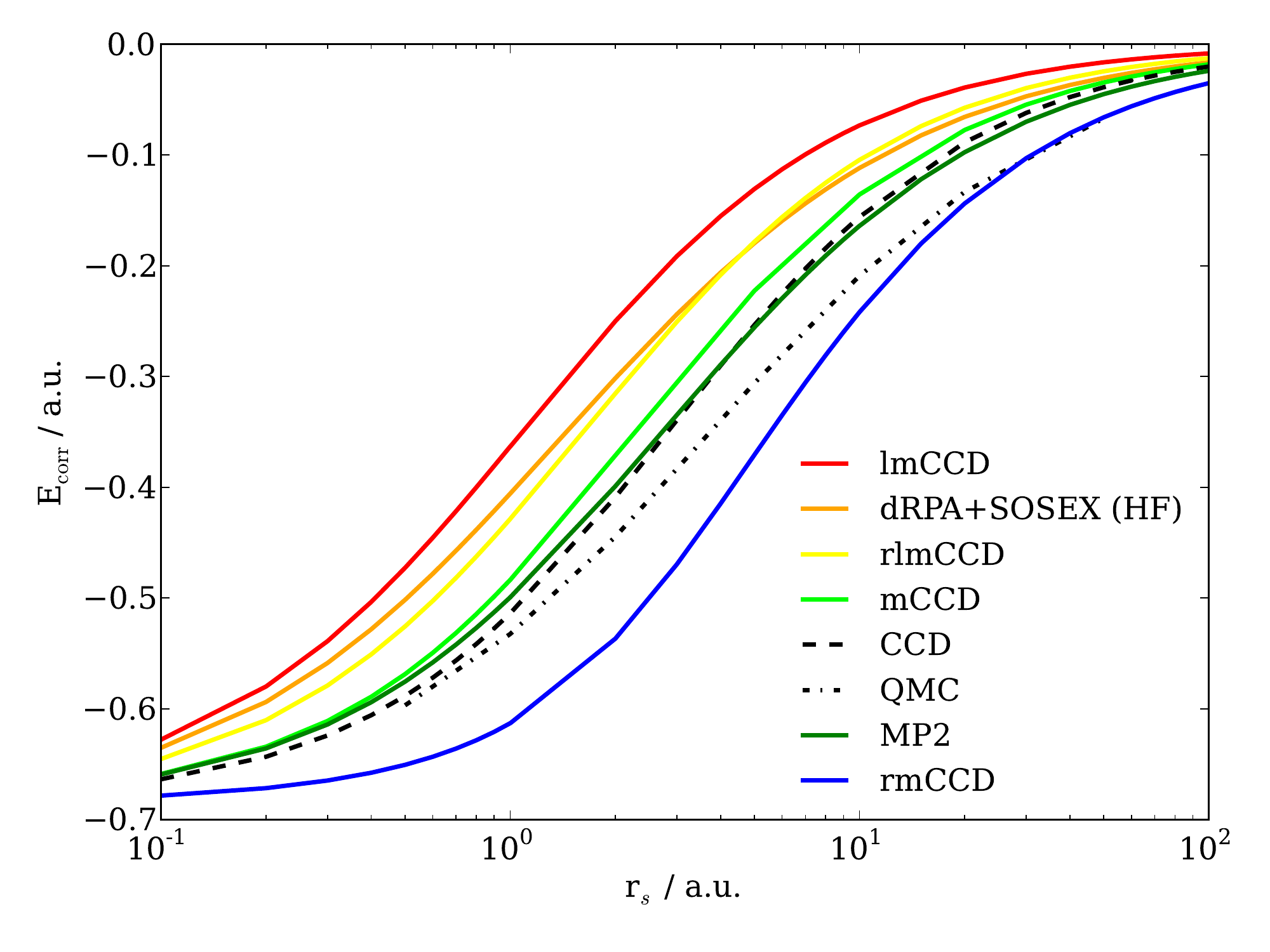}
    }
    \caption{Comparison between different types of calculations involving rings (r), crossed-rings (x), ladders (l) and mosaics (m). It can be seen that, in general, ring-based methods overestimate the amount of correlation energy. We note that, in common with figures throughout this paper, the legend is ordered top-to-bottom to indicate the ordering of lines from the left-most point of the graph to aid reading in black and white. ($N$=14, $M\rightarrow\infty$).}
    \label{fig1all}
\end{figure*}

Energies calculated using these rings-based methods are shown in \reffig{fig1}. This provides a re-iteration of some well-known general trends. Although dRPA generally overcorrelates, dRPA+SOSEX resolves this somewhat and resultantly undercorrelates. The energy from CCD is considerably improved over dRPA+SOSEX across the whole of the $r_s$ range considered when both use the same reference eigenvalues (those of HF). The dRPA+SOSEX energy is similar, for the electron gas, to CCD equations in which the rings are included only with direct integrals ($\bar{v}_{ij}^{ab}\rightarrow v_{ij}^{ab}$); the crossed-rings only with exchange integrals ($\bar{v}_{ij}^{ab}\rightarrow v_{ij}^{ba}$); and the CCD energy expression is used. These data are not shown.

In common with molecular systems, rCCD for the HEG has pathological behaviour presumably resulting from the equivalent of the Coulson--Fisher point in these systems --- the $r_s$ value where there is a symmetry-broken UHF solution. This transition has been investigated intermittently using a variety of methods~\cite{baguet_hartree-fock_2013,zhang_hartree-fock_2008,drummond_diffusion_2004}. The most recent study places the transition at densities lower than $r_s\simeq3.4$~\cite{baguet_hartree-fock_2013}. In contrast, rCCD stops converging at around $3.9>r_s>3.8$; this difference is attributed to symmetry and finite size effects. The pathology in the rCCD energy is cured by the inclusion of mosaic terms as in rmCCD, and these energies then seem reasonable at both the high and low density limits. Between these two limits, there is overcorrelation as in dRPA but less severe. The combination of rings, crossed-rings and mosaics strongly over-estimate the correlation energy and the energy curve has a spurious minimum. It is interesting to note that addition of ladders (discussed in more detail below) completely resolves this minimum and CCD is a reasonable estimate of the correlation energy.

Finally, it can also be seen from this figure that dRPA+SOSEX when determined from a Kohn--Sham reference is surprisingly accurate. This change of reference amounts to a changing of the Hartree--Fock eigenvalues to just kinetic energies (since the constant exchange-correlation term for the electron gas cancels out). This has been discussed previously and is not as generally transferable a method as CCSD~\cite{gruneis_making_2009}. The origin of this might be that on approach to the complete basis set limit the dynamic correlation is hugely overestimated especially at low densities~\cite{shepherd_convergence_2012}, compensating for the failure to capture energy from low-momentum correlations.

Energies from our calculations of CCD methods involving ladders (and remaining combinations) are shown in \reffig{fig2}. These show that in general addition of ladders cause methods to undercorrelate, with lmCCD retrieving the least correlation energy. Removing mosaics from this yields an almost identical result and is not shown. Adding ladders to rmCCD gives a result that is similar to RPA+SOSEX, and note that a screening effect from ladders was indeed anticipated by Bishop and L\"uhrman~\cite{bishop_electron_1982}. 
 
Mosaics alone give a surprisingly similar quality result to CCD, retrieving slightly less correlation energy across the whole of the energy curve. This is especially remarkable considering how much less information the mCCD equations have compared with those of CCD. On the other hand, this is perhaps related to the performance of MP2 for this system. We note, however, that the MP2 energy diverges on approach to the thermodynamic limit and this line is therefore $N$-dependent. Whether the mCCD energy diverges will be discussed in \refsec{V}.

\begin{figure}
\includegraphics[width=0.46\textwidth]{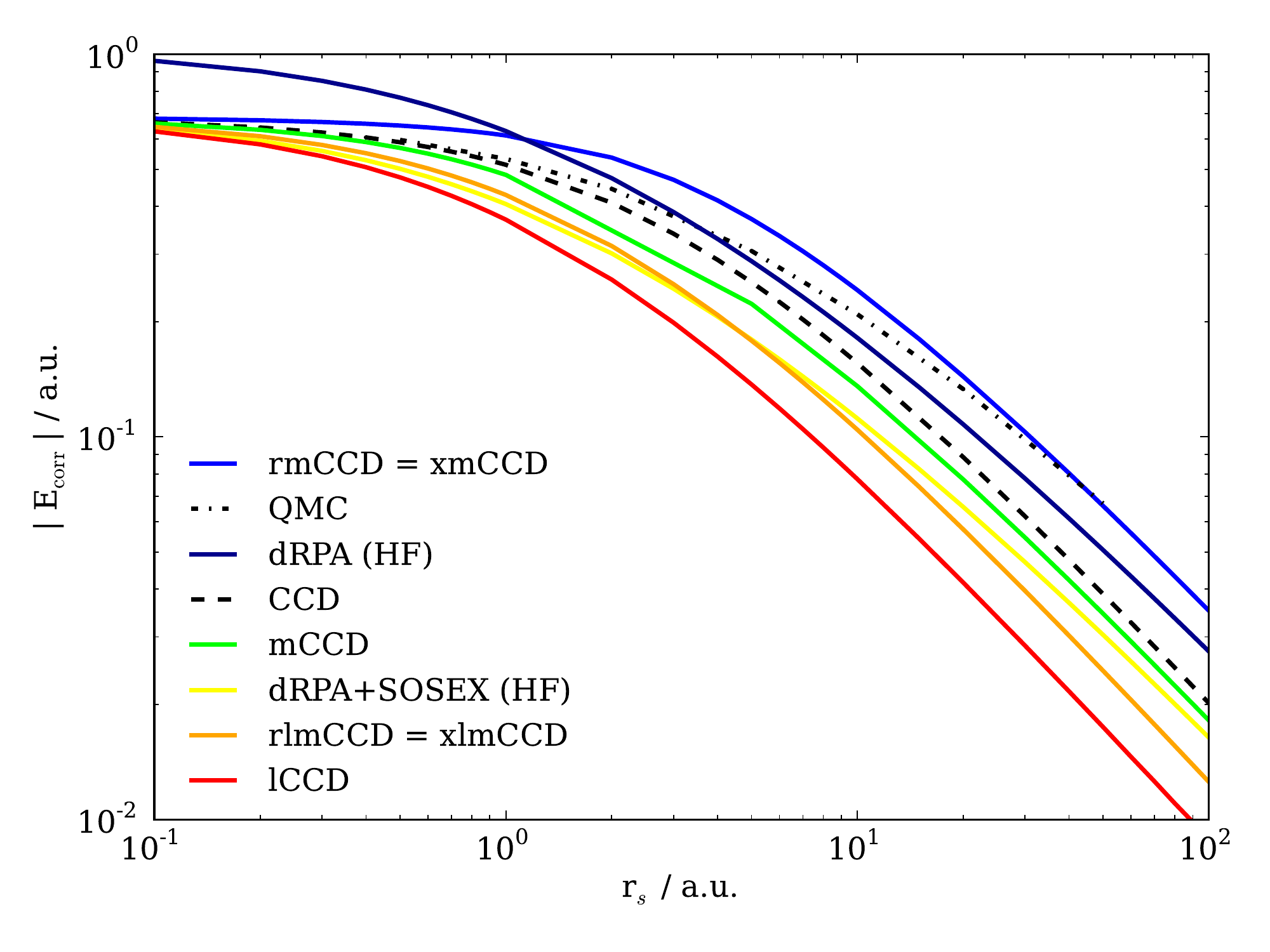}
\caption{All of the flavours of CCD have the same functional form of the low-density corresponding to 1/$r_s$ power-law behaviour. We note that the legend is ordered consistently with lines at the right-most point of the graph. ($N$=14, $M\rightarrow\infty$).}
\label{fig3}
\end{figure}

Thus, the general trend is that methods based on rings overestimate the correlation energy and ladders underestimate the correlation energy, consistent with Ref.~\cite{scuseria_particle-particle_2013}. We were not able to find any method that was a marked improvement on CCD. The low-density trend in the energy is best shown on a log-log plot of the absolute correlation energy (\reffig{fig3}), where all methods show the correct 1/$r_s$ limiting behaviour associated with approach to the cross-over to Wigner crystal behaviour. However, all of the CCDs presented here exhibit this behaviour sooner with increasing $r_s$ than the QMC results. We note in particular that ladders remain under correlated even at low densities. Similarly, most methods seem to exhibit appropriate behaviour on approach to the high-density limit $r_s\rightarrow 0$. We can see from \reffig{fig1} that the exception to this is dRPA. The reason for this is well-known, and comes from a failure of dRPA without second-order exchange to capture the constant term of the high-density expansion~\cite{gell-mann_correlation_1957,onsager_integrals_1966}. 

\section{Applicability at the thermodynamic limit}
\label{V}

The problem we wish to consider here is the approach to the infinite particle, or thermodynamic, limit (TDL) which deals with two issues: that of size extensivity and that of divergences.

Size extensivity in the current context is the requirement that as $N$ grows (whilst keeping the density constant), the energy needs to retain a term that scales as $N$. This implies that the correlation energy per particle is a non-zero constant in the TDL. This has been shown explicitly in extended systems in work by Ohnishi and Hirata, using an approach based on the linked-diagram theorem and thermodynamic scaling with volume~\cite{hirata_thermodynamic_2012}. %

Divergences are more problematic and we begin with the observation that perturbation theories applied on a bare Coulomb interaction will diverge for metallic systems in three dimensions. This arises at every order in perturbation theory (for 3D) due to the piling up of states around the Fermi energy. It is textbook knowledge that the divergence in the second-order energy (from a non-interacting starting-point) is due to a  sum over momentum transfer vectors that behaves as
$\int_{q =0} \frac{1}{q} \, \text{d} q $~\cite{mattuck_guide_1992}.

It was first shown by Gell-Mann and Brueckner that this divergence can be removed by performing a sum over the ring diagrams to infinite order, \emph{i.e.} taking the RPA rather than second-order approximations to the energy~\cite{gell-mann_correlation_1957}. Overall this yields a term in the correlation energy that behaves as $\log(r_s)$, as well as part of the constant term. The physical interpretation of this is that the interactions contributing to the diagrams become effectively screened in RPA, where they remain bare in any finite-order perturbation theory. In marked contrast, the ladders in the ladder-only approximation cannot do this, at least in the form posed by Freeman, where they have been linearized and still diverge~\cite{freeman_coupled-cluster_1983}. Instead, the ladder-only approximation is only appropriate in two dimensions, as Freeman explored~\cite{freeman_coupled-cluster_1983}, or as an addition to the RPA as achieved by Bishop and L\"uhrmann~\cite{bishop_electron_1978,bishop_electron_1982,bishop_overview_1991}. There has been some discussion of how to overcome this for three dimensions, for example by incorporation of screening effects~\cite{yasuhara_electron_1974,yasuhara_erratum:_1974,awa_electron_1980} or through use of a modified interaction~\cite{calmels_pair-correlation_1998}, although there is some suggestion that simple methods might be prone to failure~\cite{cioslowski_applicability_2005,drummond_quantum_2009}. Notwithstanding this, ladders-only theories have still seen some applications (\emph{e.g.} Ref.~\onlinecite{yurtsever_ladder_2001,kecke_ladder_2004}). 

A numerical investigation of these divergences is potentially more far-reaching in scope because the CCD equations are highly non-linear, but developments to this end have been surprisingly recent. To the best of our knowledge, the first demonstration that real fully periodic systems simulations also have divergent MP2 energies was due to Gr\"uneis and coworkers, who showed that the second-order divergence was visible in energies of sodium metal~\cite{gruneis_second-order_2010}. More recently, work by Ohnishi and Hirata take an approach inspired from thermodynamics and analyse the MP2 and CCD equations directly in terms of their scaling with volume~\cite{hirata_thermodynamic_2011,hirata_thermodynamic_2012}.

Here, we will instead follow later work, due to Shepherd and Gr\"uneis~\cite{shepherd_many-body_2013,shepherd_correlation_2012}, where it was shown that it is possible to reproduce the divergent behaviour numerically with electron gas models by simulating a series of systems of increasing electron number, $N$, and basis set size, $M$, such that $N\appropto M$ which corresponds to a ring of basis functions of fixed width around the fermi surface. The constant of proportionality does not seem particularly important, but the smallest that can easily be constructed, and hence the most efficient, is $(\sqrt{2})^3$~\footnote{For precise values of $M$ the reader is referred to Ref.~\onlinecite{shepherd_correlation_2012}}. This provides a framework for finding thermodynamic limit properties within a finite-basis method; we discuss the limitations of this approach below.

Various methods were described for how to track convergences and divergences with system size, but here we will use the MP2 energy itself as the metric for approach to the thermodynamic limit since it should diverge as $L$ to within finite-size effects. We note that methods yielding energies in proportion to the MP2 energy will also diverge. Those that do not follow the MP2 energy either converge or diverge at a slower rate. 

\begin{figure}
\includegraphics[width=0.46\textwidth]{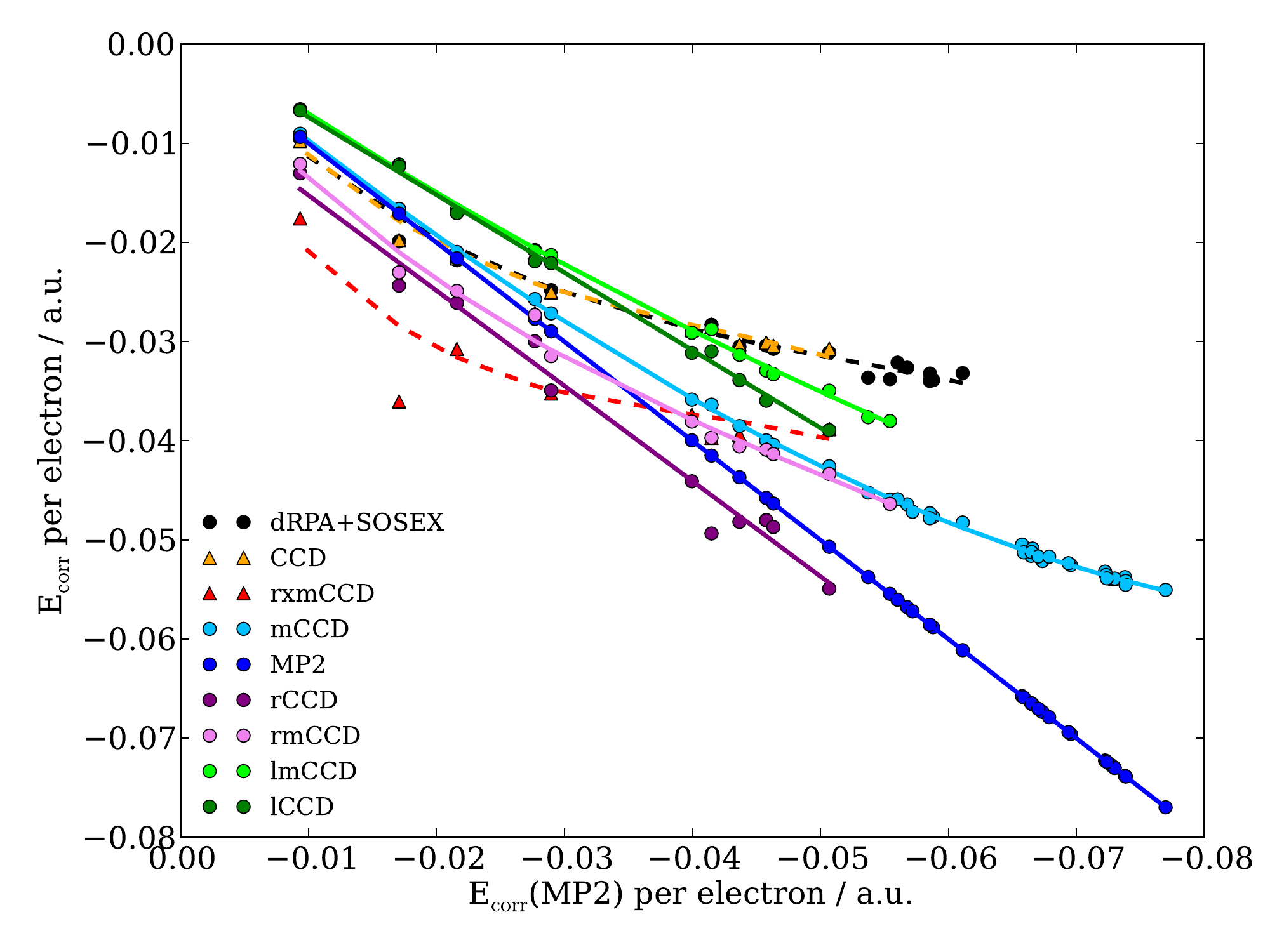}
\caption{Graph showing apparent divergences and convergences on approach to the thermodynamic limit. The method by which these are derived are discussed in the text. In particular, each point represents a system with a specific electron number (14-3006) and basis set size (M=38-8338) and the gas density is always $r_s$=1.0~a.u. This is reproduced with modifications from Ref. \onlinecite{shepherd_range_2013}.}
\label{fig7}
\end{figure}

\begin{table}
\centering
\newcolumntype{R}{>{\centering\arraybackslash}X}
\begin{tabularx}{0.48\textwidth}{R R R }
 \hline
$N$ & HF band gap (a.u.) & mCCD band gap (a.u.) \\ 
  \hline      
  \hline 
114  & -0.6950& -0.9066 \\ 
342  & -0.4332& -0.7047 \\ 
682  & -0.2927& -0.6823 \\
970  & -0.2807& -0.6365 \\ 
1598 & -0.2167& -0.6705 \\
2090 & -0.1824& -0.6931 \\
2730 & -0.1654& -0.6654 \\
3006 & -0.1507& -0.7044 \\
  \hline  
\end{tabularx}
  \caption{Band gaps for different electron numbers from Hartree--Fock theory and the renormalised Brueckner Hamiltonian one-particle eigenvalues in mCCD ($r_s=1.0$~a.u.).}  
\label{tab1}
\end{table}

\begin{figure}
\includegraphics[width=0.46\textwidth]{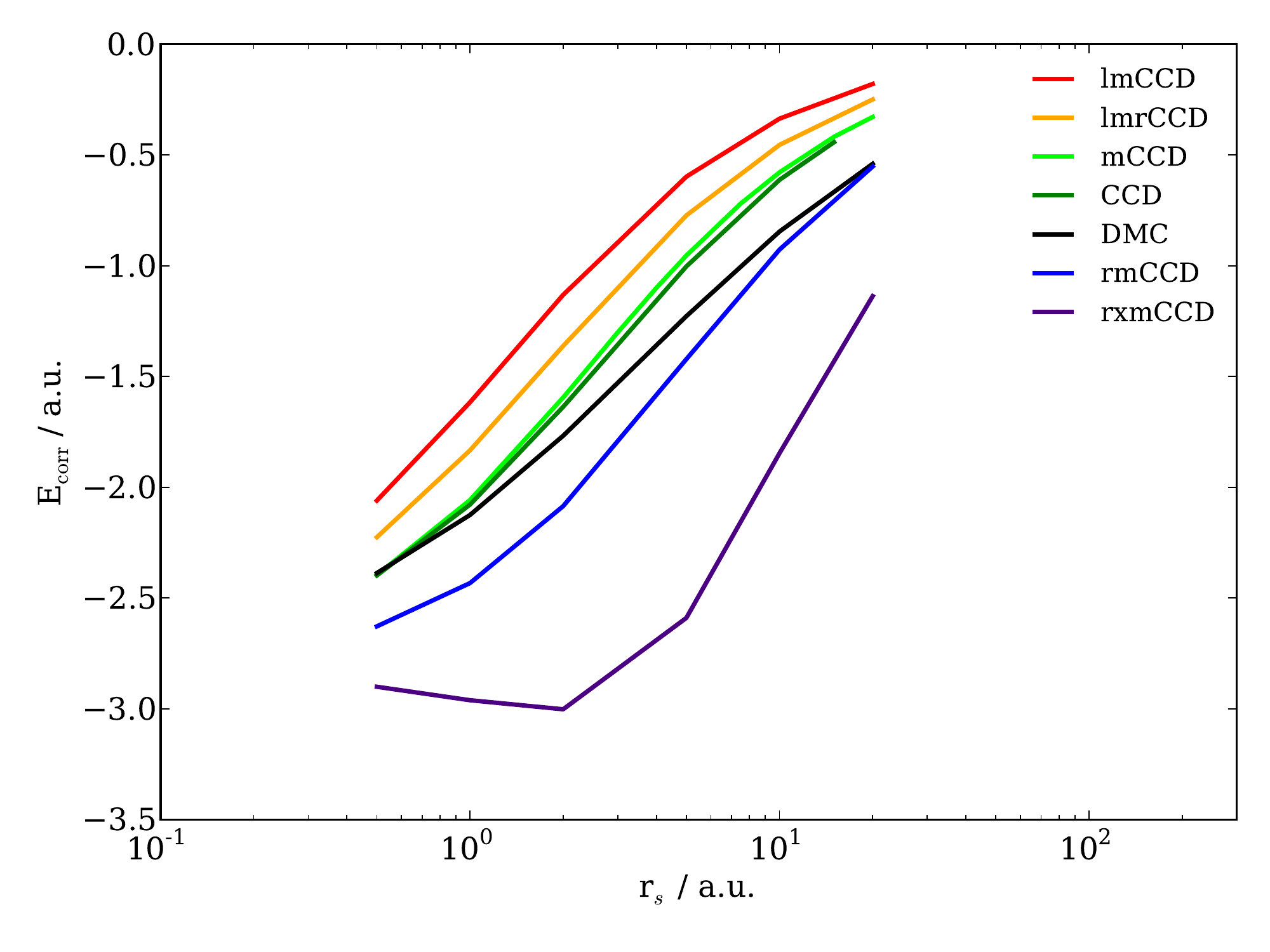} 
\caption{Comparison between different types of calculations involving rings (r), crossed-rings (x) and mosaics (m) for the 54-electron system. These show strong qualitative resemblance to the 14-electron system shown in \reffig{fig1}.}
\label{fig8}
\end{figure}

\reffig{fig7} shows the behaviour of the different channels on approach to the thermodynamic limit. We can see that the MP2 energy simply forms a diagonal line of points corresponding to different system sizes; CCD and RPA+SOSEX in contrast deviate sharply from the MP2 energy to plateau out, forming our reference for what a convergent method looks like. 

It is possible to identify from this plot three types of behaviour. The first type is those methods that simply track the MP2 energy and diverge: lCCD and rCCD. The second grouping is rxmCCD, CCD and RPA+SOSEX, methods that firmly converge. The third group are those methods that would otherwise be firmly divergent but for the presence of mosaics: mCCD, lmCCD and rmCCD. These seem to be convergent from the point of view that they drift away from their divergent counterpart (MP2, lCCD and rCCD respectively). 

The divergence of lCCD is perhaps not unexpected due to prior comments in the literature to this end~\cite{freeman_coupled-cluster_1983,bishop_electron_1978,bishop_electron_1982,bishop_overview_1991}, however, that rCCD might diverge is perhaps surprising. 
Given the equivalence of rCCD with full RPA~\cite{janesko_long-range-corrected_2009, scuseria_ground_2008,henderson_connection_2010,irelan_long-range-corrected_2011}, and the importance of full RPA in studying screening interactions~\cite{campillo_inelastic_1999,kurth_density-functional_1999,oshikiri_band_1999}, it would be worthwhile investigating this further. However, such an investigation is beyond the scope of this study.

Nevertheless, both rCCD and lCCD seem to be stabilised in the thermodynamic limit by inclusion of mosaic terms. We rationalise this behaviour of those methods by noting the mosaic terms serve to renormalise the one-electron eigenvalues. In particular for the UEG
\begin{align}
\epsilon_i^\mathrm{B} &= \epsilon_i^\mathrm{HF} + \frac{1}{2} \bar{v}_{cd}^{il}t_{il}^{cd}  \\
{\epsilon}_a^\mathrm{B} &= \epsilon_a^\mathrm{HF} - \frac{1}{2} \bar{v}_{ad}^{kl}  t_{kl}^{ad}  
\end{align}
for occupied $i$ and virtual $a$. These in general must be solved self-consistently with the corresponding amplitude equations. Thus, mCCD is a kind of self-consistent second-order perturbation theory from a Brueckner one-body Hamiltonian reference~\cite{nesbet_brueckners_1958} where $t=v/\Delta\tilde{\epsilon}$ as discussed earlier. Because $t<0$ and $v>0$~\footnote{This has been investigated explicitly by looking at the exact wavefunction in Ref.~\onlinecite{shepherd_quantum_2013}}, this renormalisation serves to open a gap for the HEG. \reftab{tab1} demonstrates that for a series of systems up to $N=3006$, the band gap for the renormalised eigenvalues remains open. This implies that all resultant theories are performed with respect to an insulating reference and should be expected to converge, and this explains the behaviour shown in \reffig{fig7}. To confirm this, we artificially gapped the Hartree--Fock eigenspectrum with a constant gap and this shows remarkably similar behaviour to the mCCD on approach to the thermodynamic limit. This is consistent with the removal of the divergence of the form 
$\int_{q =0} \frac{1}{q+\xi} \, \text{d} q, \xi>0.$
It is unclear as to whether this, while being beneficial to the convergence of a second-order energy, has any physical interpretation.

Much of what we described for the 14 electron system holds at this stage in our discussion, in spite of the discovery of some divergences. For the rings-only CCD, mosaics were required to give a meaningful energy curve free of pathologies. For the ladders, addition of mosaics made very little energetic difference. As far as generalising the conclusions we drew to other electron numbers, it is possible to see that ladders are still the most undercorrelating of all the (mosaic-based) methods, although rmCCD seems to end up somewhere closer to mCCD for this basis set size. The same qualitative features of the 14 electron system are also present in the 54 electron system (\reffig{fig8}). This mirrors results from molecular systems~\cite{scuseria_particle-particle_2013}; it remains to be seen if these results are transferable to other extended systems.

Another question we need to address is the strength of this analysis. One of the limitations that this has is that it can only examine divergences or convergences of a certain length scale --- here we choose the MP2 energy divergence and the RPA energy convergence as our exemplars. If the energy of a method diverges or converges on a different length scale to either MP2 or RPA, this test would almost certainly fail to notice this. There is also the effect of high-lying parts of the momentum space which we have omitted from our results: Could these somehow effect the calculation? There is no available answer to this other than to remark that analytically predicted trends are reproduced which lend credence to these numerical findings. Perhaps one of the troublesome aspects of this analysis is that from \reffig{fig7}, it would seem that mosaics correlate at this basis set size very strongly --- far more strongly than RPA. Even if the method turns around as does mCCD, is it just going to be wildly over-correlated? This is unclear, as we can only approach the thermodynamic limit whilst also being far from the complete basis set limit; we shall leave this as an open question at this time.

\section{Conclusions}

In conclusion we have used modified and approximate coupled cluster doubles (CCD) equations, where terms have been separated into different classes of diagrams, to study the energies of finite electron gases ($N=14$ and 54) for a wide variety of densities and have compared our findings to modern quantum Monte Carlo data. We find in general that no combination of channels performs as well as CCD in reproducing energies. In particular, rings-based approximations tend to overestimate correlation energies and methods involving ladders underestimate correlation energies. 

Approaching the thermodynamic limit, we present numerical findings suggesting that rings-only CCD and ladders-only CCD diverge at a similar rate to the second-order energy (taken here from MP2). The rings-only divergence can be cured either by removing the antisymmetrization on the four-index integrals to make direct RPA, or by inclusion of crossed-rings. We present strong evidence that inclusion of mosaic terms self-consistently renormalises the Hartree--Fock eigenvalues, opening a band gap and making methods including mosaic terms converge at the thermodynamic limit. These include mosaic-only CCD, which is equivalent to self-consistent MP2 based on the Brueckner one-body Hamiltonian.

\medskip

\acknowledgements 

The authors thank Andreas Gr\"uneis for helpful discussions and the authors of Ref.~\onlinecite{shepherd__2013} for early access to their data. One of us (JJS) would like to thank the Royal Commission for the Exhibition of 1851 for a Research Fellowship (2013--2016). This work was supported by the National Science Foundation (CHE-1102601), the Welch Foundation (C-0036) and DOE-CMCSN (DE-SC0006650).

\begin{table*}
\centering
\newcolumntype{R}{>{\centering\arraybackslash}X}
\begin{tabularx}{1.0\textwidth}{ R R R R R R R }
 \hline
$r_s$ (a.u.) & lmCCD (a.u.)  & lmrCCD (a.u.)  & mCCD (a.u.)  & CCD (a.u.)  & rmCCD (a.u.)  & rxmCCD (a.u.)  \\ 
  \hline      
  \hline 
  0.5 & -2.038 & -2.184 & -2.394 & -2.372 & -2.620 & -2.885  \\ 
1.0 & -1.600 & -1.799 & -2.055 & -2.052 & -2.423 & -2.941  \\ 
2.0 & -1.119 & -1.331 & -1.592 & -1.620 & -2.077 & -2.977  \\ 
3.0 & - & - & -1.299 & - & - & -  \\ 
4.0 & - & - & -1.098 & - & - & -  \\ 
5.0 & -0.591 & -0.756 & -0.952 & -0.995 & -1.413 & -2.576  \\ 
7.5 & - & - & -0.717 & - & - & -  \\ 
10.0 & -0.331 & -0.440 & -0.577 & -0.609 & -0.919 & -1.839  \\ 
15.0 & - & - & -0.416 & -0.440 & - & -  \\ 
20.0 & -0.176 & -0.242 & -0.327 & - & -0.545 & -1.131  \\ 
  \hline  
\end{tabularx}
  \caption{Complete basis set energies for the 54 electron gas. Abbreviations are explained in the accompanying paper, plotted in \reffig{fig8}.}
\label{54ers1table}
\end{table*}

\begin{table*}
\centering
\newcolumntype{R}{>{\centering\arraybackslash}X}
\begin{tabularx}{1.0\textwidth}{ R R R R R R R R R R }
 \hline
$r_s$ (a.u.) & lmCCD (a.u.)  & mCCD (a.u.)  & CCD (a.u.)  & rmCCD (a.u.)  & rxmCCD (a.u.)  & d+S (KS) (a.u.)  & dRPA (KS) (a.u.)  & d+S (HF) (a.u.)  & dRPA (HF) (a.u.)  \\ 
 \hline
  \hline
0.1 & -0.627 & -0.659 & -0.663 & -0.678 & -0.699 & -0.658 & -0.995 & -0.635 & -0.961  \\ 
0.2 & -0.579 & -0.634 & -0.643 & -0.672 & -0.713 & -0.635 & -0.963 & -0.594 & -0.902  \\ 
0.3 & -0.538 & -0.611 & -0.624 & -0.665 & -0.728 & -0.614 & -0.935 & -0.559 & -0.852  \\ 
0.4 & -0.503 & -0.589 & -0.605 & -0.658 & -0.743 & -0.595 & -0.910 & -0.528 & -0.808  \\ 
0.5 & -0.472 & -0.569 & -0.588 & -0.651 & -0.758 & -0.577 & -0.886 & -0.502 & -0.769  \\ 
0.6 & -0.445 & -0.549 & -0.572 & -0.643 & -0.774 & -0.561 & -0.865 & -0.478 & -0.735  \\ 
0.7 & -0.421 & -0.531 & -0.556 & -0.636 & -0.790 & -0.547 & -0.845 & -0.457 & -0.705  \\ 
0.8 & -0.400 & -0.515 & -0.541 & -0.628 & -0.807 & -0.533 & -0.827 & -0.438 & -0.677  \\ 
0.9 & -0.380 & -0.499 & -0.527 & -0.621 & -0.824 & -0.521 & -0.809 & -0.421 & -0.652  \\ 
1.0 & -0.363 & -0.484 & -0.514 & -0.613 & -0.841 & -0.509 & -0.794 & -0.406 & -0.629  \\ 
2.0 & -0.250 & - & -0.409 & -0.537 & -0.993 & -0.423 & -0.673 & -0.301 & -0.474  \\ 
3.0 & -0.191 & - & -0.339 & -0.469 & -0.994 & -0.368 & -0.596 & -0.244 & -0.387  \\ 
4.0 & -0.155 & - & -0.290 & -0.415 & -0.917 & -0.330 & -0.541 & -0.206 & -0.329  \\ 
5.0 & -0.131 & -0.223 & -0.253 & -0.371 & -0.833 & -0.301 & -0.499 & -0.180 & -0.288  \\ 
6.0 & -0.113 & - & -0.225 & -0.335 & -0.756 & -0.278 & -0.464 & -0.159 & -0.256  \\ 
7.0 & -0.099 & - & -0.203 & -0.305 & -0.691 & -0.259 & -0.436 & -0.144 & -0.232  \\ 
8.0 & -0.089 & - & -0.184 & -0.281 & -0.634 & -0.243 & -0.412 & -0.131 & -0.212  \\ 
9.0 & -0.080 & - & -0.169 & -0.260 & -0.586 & -0.230 & -0.391 & -0.121 & -0.195  \\ 
10.0 & -0.073 & -0.136 & -0.156 & -0.242 & -0.545 & -0.218 & -0.373 & -0.112 & -0.181  \\ 
15.0 & -0.051 & - & - & -0.180 & -0.402 & - & - & -0.082 & -0.135  \\ 
20.0 & -0.039 & -0.077 & -0.089 & -0.144 & -0.319 & - & - & -0.066 & -0.108  \\ 
30.0 & -0.027 & -0.055 & -0.062 & -0.103 & -0.227 & - & - & -0.047 & -0.078  \\ 
40.0 & -0.020 & -0.042 & -0.048 & -0.080 & -0.176 & - & - & -0.037 & -0.061  \\ 
50.0 & -0.016 & -0.034 & -0.039 & -0.066 & -0.144 & - & - & -0.030 & -0.051  \\ 
60.0 & -0.014 & -0.029 & -0.033 & -0.056 & -0.122 & - & - & -0.026 & -0.043  \\ 
70.0 & -0.012 & -0.025 & -0.028 & -0.049 & -0.106 & - & - & -0.023 & -0.038  \\ 
80.0 & -0.010 & -0.022 & -0.025 & -0.043 & -0.094 & - & - & -0.020 & -0.034  \\ 
90.0 & -0.009 & -0.020 & -0.022 & -0.039 & -0.084 & - & - & -0.018 & -0.030  \\ 
100.0 & -0.008 & -0.018 & -0.020 & -0.035 & -0.076 & - & - & -0.016 & -0.027  \\ 
  \hline  
\end{tabularx}
  \caption{Complete basis set energies for the 14 electron gas. Abbreviations are explained in the accompanying paper, plotted in \reffig{fig3}. ``d+S'' refers to dRPA+SOSEX.}
\label{54ers1table}
\end{table*}

\clearpage


\begin{thebibliography}{139}%
\makeatletter
\providecommand \@ifxundefined [1]{%
 \@ifx{#1\undefined}
}%
\providecommand \@ifnum [1]{%
 \ifnum #1\expandafter \@firstoftwo
 \else \expandafter \@secondoftwo
 \fi
}%
\providecommand \@ifx [1]{%
 \ifx #1\expandafter \@firstoftwo
 \else \expandafter \@secondoftwo
 \fi
}%
\providecommand \natexlab [1]{#1}%
\providecommand \enquote  [1]{``#1''}%
\providecommand \bibnamefont  [1]{#1}%
\providecommand \bibfnamefont [1]{#1}%
\providecommand \citenamefont [1]{#1}%
\providecommand \href@noop [0]{\@secondoftwo}%
\providecommand \href [0]{\begingroup \@sanitize@url \@href}%
\providecommand \@href[1]{\@@startlink{#1}\@@href}%
\providecommand \@@href[1]{\endgroup#1\@@endlink}%
\providecommand \@sanitize@url [0]{\catcode `\\12\catcode `\$12\catcode
  `\&12\catcode `\#12\catcode `\^12\catcode `\_12\catcode `\%12\relax}%
\providecommand \@@startlink[1]{}%
\providecommand \@@endlink[0]{}%
\providecommand \url  [0]{\begingroup\@sanitize@url \@url }%
\providecommand \@url [1]{\endgroup\@href {#1}{\urlprefix }}%
\providecommand \urlprefix  [0]{URL }%
\providecommand \Eprint [0]{\href }%
\providecommand \doibase [0]{http://dx.doi.org/}%
\providecommand \selectlanguage [0]{\@gobble}%
\providecommand \bibinfo  [0]{\@secondoftwo}%
\providecommand \bibfield  [0]{\@secondoftwo}%
\providecommand \translation [1]{[#1]}%
\providecommand \BibitemOpen [0]{}%
\providecommand \bibitemStop [0]{}%
\providecommand \bibitemNoStop [0]{.\EOS\space}%
\providecommand \EOS [0]{\spacefactor3000\relax}%
\providecommand \BibitemShut  [1]{\csname bibitem#1\endcsname}%
\let\auto@bib@innerbib\@empty
\bibitem [{\citenamefont {Scuseria}\ \emph {et~al.}(2008)\citenamefont
  {Scuseria}, \citenamefont {Henderson},\ and\ \citenamefont
  {Sorensen}}]{scuseria_ground_2008}%
  \BibitemOpen
  \bibfield  {author} {\bibinfo {author} {\bibfnamefont {G.~E.}\ \bibnamefont
  {Scuseria}}, \bibinfo {author} {\bibfnamefont {T.~M.}\ \bibnamefont
  {Henderson}}, \ and\ \bibinfo {author} {\bibfnamefont {D.~C.}\ \bibnamefont
  {Sorensen}},\ }\href {\doibase doi:10.1063/1.3043729} {\bibfield  {journal}
  {\bibinfo  {journal} {J. Chem. Phys.}\ }\textbf {\bibinfo {volume} {129}},\
  \bibinfo {pages} {231101} (\bibinfo {year} {2008})}\BibitemShut {NoStop}%
\bibitem [{\citenamefont {Scuseria}\ \emph {et~al.}(2013)\citenamefont
  {Scuseria}, \citenamefont {Henderson},\ and\ \citenamefont
  {Bulik}}]{scuseria_particle-particle_2013}%
  \BibitemOpen
  \bibfield  {author} {\bibinfo {author} {\bibfnamefont {G.~E.}\ \bibnamefont
  {Scuseria}}, \bibinfo {author} {\bibfnamefont {T.~M.}\ \bibnamefont
  {Henderson}}, \ and\ \bibinfo {author} {\bibfnamefont {I.~W.}\ \bibnamefont
  {Bulik}},\ }\href {\doibase doi:10.1063/1.4820557} {\bibfield  {journal}
  {\bibinfo  {journal} {J. Chem. Phys.}\ }\textbf {\bibinfo {volume} {139}},\
  \bibinfo {pages} {104113} (\bibinfo {year} {2013})}\BibitemShut {NoStop}%
\bibitem [{\citenamefont {Peng}\ \emph {et~al.}(2013)\citenamefont {Peng},
  \citenamefont {Steinmann}, \citenamefont {van Aggelen},\ and\ \citenamefont
  {Yang}}]{peng_equivalence_2013}%
  \BibitemOpen
  \bibfield  {author} {\bibinfo {author} {\bibfnamefont {D.}~\bibnamefont
  {Peng}}, \bibinfo {author} {\bibfnamefont {S.~N.}\ \bibnamefont {Steinmann}},
  \bibinfo {author} {\bibfnamefont {H.}~\bibnamefont {van Aggelen}}, \ and\
  \bibinfo {author} {\bibfnamefont {W.}~\bibnamefont {Yang}},\ }\href {\doibase
  doi:10.1063/1.4820556} {\bibfield  {journal} {\bibinfo  {journal} {J. Chem.
  Phys.}\ }\textbf {\bibinfo {volume} {139}},\ \bibinfo {pages} {104112}
  (\bibinfo {year} {2013})}\BibitemShut {NoStop}%
\bibitem [{\citenamefont {van Aggelen}\ \emph {et~al.}(2013)\citenamefont {van
  Aggelen}, \citenamefont {Yang},\ and\ \citenamefont
  {Yang}}]{van_aggelen_exchange-correlation_2013}%
  \BibitemOpen
  \bibfield  {author} {\bibinfo {author} {\bibfnamefont {H.}~\bibnamefont {van
  Aggelen}}, \bibinfo {author} {\bibfnamefont {Y.}~\bibnamefont {Yang}}, \ and\
  \bibinfo {author} {\bibfnamefont {W.}~\bibnamefont {Yang}},\ }\href {\doibase
  10.1103/PhysRevA.88.030501} {\bibfield  {journal} {\bibinfo  {journal} {Phys.
  Rev. A}\ }\textbf {\bibinfo {volume} {88}},\ \bibinfo {pages} {030501}
  (\bibinfo {year} {2013})}\BibitemShut {NoStop}%
\bibitem [{\citenamefont {{\v C}{\'i}{\v z}ek}(1966)}]{cizek_correlation_1966}%
  \BibitemOpen
  \bibfield  {author} {\bibinfo {author} {\bibfnamefont {J.}~\bibnamefont {{\v
  C}{\'i}{\v z}ek}},\ }\href {\doibase 10.1063/1.1727484} {\bibfield  {journal}
  {\bibinfo  {journal} {J. Chem. Phys.}\ }\textbf {\bibinfo {volume} {45}},\
  \bibinfo {pages} {4256} (\bibinfo {year} {1966})}\BibitemShut {NoStop}%
\bibitem [{\citenamefont {Shepherd}\ \emph
  {et~al.}(2012{\natexlab{a}})\citenamefont {Shepherd}, \citenamefont {Booth},
  \citenamefont {Gr{\"u}neis},\ and\ \citenamefont
  {Alavi}}]{shepherd_full_2012}%
  \BibitemOpen
  \bibfield  {author} {\bibinfo {author} {\bibfnamefont {J.~J.}\ \bibnamefont
  {Shepherd}}, \bibinfo {author} {\bibfnamefont {G.}~\bibnamefont {Booth}},
  \bibinfo {author} {\bibfnamefont {A.}~\bibnamefont {Gr{\"u}neis}}, \ and\
  \bibinfo {author} {\bibfnamefont {A.}~\bibnamefont {Alavi}},\ }\href
  {\doibase 10.1103/PhysRevB.85.081103} {\bibfield  {journal} {\bibinfo
  {journal} {Phys. Rev. B}\ }\textbf {\bibinfo {volume} {85}},\ \bibinfo
  {pages} {081103} (\bibinfo {year} {2012}{\natexlab{a}})}\BibitemShut
  {NoStop}%
\bibitem [{\citenamefont {Shepherd}\ \emph
  {et~al.}(2012{\natexlab{b}})\citenamefont {Shepherd}, \citenamefont {Booth},\
  and\ \citenamefont {Alavi}}]{shepherd_investigation_2012}%
  \BibitemOpen
  \bibfield  {author} {\bibinfo {author} {\bibfnamefont {J.~J.}\ \bibnamefont
  {Shepherd}}, \bibinfo {author} {\bibfnamefont {G.~H.}\ \bibnamefont {Booth}},
  \ and\ \bibinfo {author} {\bibfnamefont {A.}~\bibnamefont {Alavi}},\ }\href
  {\doibase 10.1063/1.4720076} {\bibfield  {journal} {\bibinfo  {journal} {J.
  Chem. Phys.}\ }\textbf {\bibinfo {volume} {136}},\ \bibinfo {pages} {244101}
  (\bibinfo {year} {2012}{\natexlab{b}})}\BibitemShut {NoStop}%
\bibitem [{\citenamefont {Shepherd}\ \emph
  {et~al.}(2012{\natexlab{c}})\citenamefont {Shepherd}, \citenamefont
  {Gr{\"u}neis}, \citenamefont {Booth}, \citenamefont {Kresse},\ and\
  \citenamefont {Alavi}}]{shepherd_convergence_2012}%
  \BibitemOpen
  \bibfield  {author} {\bibinfo {author} {\bibfnamefont {J.~J.}\ \bibnamefont
  {Shepherd}}, \bibinfo {author} {\bibfnamefont {A.}~\bibnamefont
  {Gr{\"u}neis}}, \bibinfo {author} {\bibfnamefont {G.~H.}\ \bibnamefont
  {Booth}}, \bibinfo {author} {\bibfnamefont {G.}~\bibnamefont {Kresse}}, \
  and\ \bibinfo {author} {\bibfnamefont {A.}~\bibnamefont {Alavi}},\ }\href
  {\doibase 10.1103/PhysRevB.86.035111} {\bibfield  {journal} {\bibinfo
  {journal} {Phys. Rev. B}\ }\textbf {\bibinfo {volume} {86}},\ \bibinfo
  {pages} {035111} (\bibinfo {year} {2012}{\natexlab{c}})}\BibitemShut
  {NoStop}%
\bibitem [{\citenamefont {Shepherd}\ and\ \citenamefont
  {Gr{\"u}neis}(2013)}]{shepherd_many-body_2013}%
  \BibitemOpen
  \bibfield  {author} {\bibinfo {author} {\bibfnamefont {J.~J.}\ \bibnamefont
  {Shepherd}}\ and\ \bibinfo {author} {\bibfnamefont {A.}~\bibnamefont
  {Gr{\"u}neis}},\ }\href {\doibase 10.1103/PhysRevLett.110.226401} {\bibfield
  {journal} {\bibinfo  {journal} {Phys. Rev. Lett.}\ }\textbf {\bibinfo
  {volume} {110}},\ \bibinfo {pages} {226401} (\bibinfo {year}
  {2013})}\BibitemShut {NoStop}%
\bibitem [{\citenamefont {Gr{\"u}neis}\ \emph {et~al.}(2013)\citenamefont
  {Gr{\"u}neis}, \citenamefont {Shepherd}, \citenamefont {Alavi}, \citenamefont
  {Tew},\ and\ \citenamefont {Booth}}]{gruneis_explicitly_2013}%
  \BibitemOpen
  \bibfield  {author} {\bibinfo {author} {\bibfnamefont {A.}~\bibnamefont
  {Gr{\"u}neis}}, \bibinfo {author} {\bibfnamefont {J.~J.}\ \bibnamefont
  {Shepherd}}, \bibinfo {author} {\bibfnamefont {A.}~\bibnamefont {Alavi}},
  \bibinfo {author} {\bibfnamefont {D.~P.}\ \bibnamefont {Tew}}, \ and\
  \bibinfo {author} {\bibfnamefont {G.~H.}\ \bibnamefont {Booth}},\ }\href
  {\doibase doi:10.1063/1.4818753} {\bibfield  {journal} {\bibinfo  {journal}
  {J. Chem. Phys.}\ }\textbf {\bibinfo {volume} {139}},\ \bibinfo {pages}
  {084112} (\bibinfo {year} {2013})}\BibitemShut {NoStop}%
\bibitem [{\citenamefont {Marsman}\ \emph {et~al.}(2009)\citenamefont
  {Marsman}, \citenamefont {Gr{\"u}neis}, \citenamefont {Paier},\ and\
  \citenamefont {Kresse}}]{marsman_second-order_2009}%
  \BibitemOpen
  \bibfield  {author} {\bibinfo {author} {\bibfnamefont {M.}~\bibnamefont
  {Marsman}}, \bibinfo {author} {\bibfnamefont {A.}~\bibnamefont
  {Gr{\"u}neis}}, \bibinfo {author} {\bibfnamefont {J.}~\bibnamefont {Paier}},
  \ and\ \bibinfo {author} {\bibfnamefont {G.}~\bibnamefont {Kresse}},\ }\href
  {\doibase 10.1063/1.3126249} {\bibfield  {journal} {\bibinfo  {journal} {J.
  Chem. Phys.}\ }\textbf {\bibinfo {volume} {130}},\ \bibinfo {pages} {184103}
  (\bibinfo {year} {2009})}\BibitemShut {NoStop}%
\bibitem [{\citenamefont {L{\'o}pez~R{\'i}os}\ \emph
  {et~al.}(2006)\citenamefont {L{\'o}pez~R{\'i}os}, \citenamefont {Ma},
  \citenamefont {Drummond}, \citenamefont {Towler},\ and\ \citenamefont
  {Needs}}]{lopez_rios_inhomogeneous_2006}%
  \BibitemOpen
  \bibfield  {author} {\bibinfo {author} {\bibfnamefont {P.}~\bibnamefont
  {L{\'o}pez~R{\'i}os}}, \bibinfo {author} {\bibfnamefont {A.}~\bibnamefont
  {Ma}}, \bibinfo {author} {\bibfnamefont {N.~D.}\ \bibnamefont {Drummond}},
  \bibinfo {author} {\bibfnamefont {M.~D.}\ \bibnamefont {Towler}}, \ and\
  \bibinfo {author} {\bibfnamefont {R.~J.}\ \bibnamefont {Needs}},\ }\href
  {\doibase 10.1103/PhysRevE.74.066701} {\bibfield  {journal} {\bibinfo
  {journal} {Phys. Rev. E}\ }\textbf {\bibinfo {volume} {74}},\ \bibinfo
  {pages} {066701} (\bibinfo {year} {2006})}\BibitemShut {NoStop}%
\bibitem [{\citenamefont {Kwon}\ \emph {et~al.}(1998)\citenamefont {Kwon},
  \citenamefont {Ceperley},\ and\ \citenamefont {Martin}}]{kwon_effects_1998}%
  \BibitemOpen
  \bibfield  {author} {\bibinfo {author} {\bibfnamefont {Y.}~\bibnamefont
  {Kwon}}, \bibinfo {author} {\bibfnamefont {D.~M.}\ \bibnamefont {Ceperley}},
  \ and\ \bibinfo {author} {\bibfnamefont {R.~M.}\ \bibnamefont {Martin}},\
  }\href {\doibase 10.1103/PhysRevB.58.6800} {\bibfield  {journal} {\bibinfo
  {journal} {Phys. Rev. B}\ }\textbf {\bibinfo {volume} {58}},\ \bibinfo
  {pages} {6800} (\bibinfo {year} {1998})}\BibitemShut {NoStop}%
\bibitem [{\citenamefont {Holzmann}\ \emph {et~al.}(2003)\citenamefont
  {Holzmann}, \citenamefont {Ceperley}, \citenamefont {Pierleoni},\ and\
  \citenamefont {Esler}}]{holzmann_backflow_2003}%
  \BibitemOpen
  \bibfield  {author} {\bibinfo {author} {\bibfnamefont {M.}~\bibnamefont
  {Holzmann}}, \bibinfo {author} {\bibfnamefont {D.~M.}\ \bibnamefont
  {Ceperley}}, \bibinfo {author} {\bibfnamefont {C.}~\bibnamefont {Pierleoni}},
  \ and\ \bibinfo {author} {\bibfnamefont {K.}~\bibnamefont {Esler}},\ }\href
  {\doibase 10.1103/PhysRevE.68.046707} {\bibfield  {journal} {\bibinfo
  {journal} {Phys. Rev. E}\ }\textbf {\bibinfo {volume} {68}},\ \bibinfo
  {pages} {046707} (\bibinfo {year} {2003})}\BibitemShut {NoStop}%
\bibitem [{\citenamefont {Ortiz}\ and\ \citenamefont
  {Ballone}(1994)}]{ortiz_correlation_1994}%
  \BibitemOpen
  \bibfield  {author} {\bibinfo {author} {\bibfnamefont {G.}~\bibnamefont
  {Ortiz}}\ and\ \bibinfo {author} {\bibfnamefont {P.}~\bibnamefont
  {Ballone}},\ }\href {\doibase 10.1103/PhysRevB.50.1391} {\bibfield  {journal}
  {\bibinfo  {journal} {Phys. Rev. B}\ }\textbf {\bibinfo {volume} {50}},\
  \bibinfo {pages} {1391} (\bibinfo {year} {1994})}\BibitemShut {NoStop}%
\bibitem [{\citenamefont {Ceperley}\ and\ \citenamefont
  {Alder}(1980)}]{ceperley_ground_1980}%
  \BibitemOpen
  \bibfield  {author} {\bibinfo {author} {\bibfnamefont {D.~M.}\ \bibnamefont
  {Ceperley}}\ and\ \bibinfo {author} {\bibfnamefont {B.~J.}\ \bibnamefont
  {Alder}},\ }\href {\doibase 10.1103/PhysRevLett.45.566} {\bibfield  {journal}
  {\bibinfo  {journal} {Phys. Rev. Lett.}\ }\textbf {\bibinfo {volume} {45}},\
  \bibinfo {pages} {566} (\bibinfo {year} {1980})}\BibitemShut {NoStop}%
\bibitem [{\citenamefont {Spink}\ \emph {et~al.}(2013)\citenamefont {Spink},
  \citenamefont {Needs},\ and\ \citenamefont {Drummond}}]{spink_quantum_2013}%
  \BibitemOpen
  \bibfield  {author} {\bibinfo {author} {\bibfnamefont {G.~G.}\ \bibnamefont
  {Spink}}, \bibinfo {author} {\bibfnamefont {R.~J.}\ \bibnamefont {Needs}}, \
  and\ \bibinfo {author} {\bibfnamefont {N.~D.}\ \bibnamefont {Drummond}},\
  }\href {\doibase 10.1103/PhysRevB.88.085121} {\bibfield  {journal} {\bibinfo
  {journal} {Phys. Rev. B}\ }\textbf {\bibinfo {volume} {88}},\ \bibinfo
  {pages} {085121} (\bibinfo {year} {2013})}\BibitemShut {NoStop}%
\bibitem [{\citenamefont {Drummond}\ and\ \citenamefont
  {Needs}(2013{\natexlab{a}})}]{drummond_quantum_2013}%
  \BibitemOpen
  \bibfield  {author} {\bibinfo {author} {\bibfnamefont {N.~D.}\ \bibnamefont
  {Drummond}}\ and\ \bibinfo {author} {\bibfnamefont {R.~J.}\ \bibnamefont
  {Needs}},\ }\href {\doibase 10.1103/PhysRevB.88.035133} {\bibfield  {journal}
  {\bibinfo  {journal} {Phys. Rev. B}\ }\textbf {\bibinfo {volume} {88}},\
  \bibinfo {pages} {035133} (\bibinfo {year} {2013}{\natexlab{a}})}\BibitemShut
  {NoStop}%
\bibitem [{\citenamefont {Drummond}\ and\ \citenamefont
  {Needs}(2013{\natexlab{b}})}]{drummond_diffusion_2013}%
  \BibitemOpen
  \bibfield  {author} {\bibinfo {author} {\bibfnamefont {N.~D.}\ \bibnamefont
  {Drummond}}\ and\ \bibinfo {author} {\bibfnamefont {R.~J.}\ \bibnamefont
  {Needs}},\ }\href {\doibase 10.1103/PhysRevB.87.045131} {\bibfield  {journal}
  {\bibinfo  {journal} {Phys. Rev. B}\ }\textbf {\bibinfo {volume} {87}},\
  \bibinfo {pages} {045131} (\bibinfo {year} {2013}{\natexlab{b}})}\BibitemShut
  {NoStop}%
\bibitem [{\citenamefont {Brown}\ \emph {et~al.}(2013)\citenamefont {Brown},
  \citenamefont {{DuBois}}, \citenamefont {Holzmann},\ and\ \citenamefont
  {Ceperley}}]{brown_exchange-correlation_2013}%
  \BibitemOpen
  \bibfield  {author} {\bibinfo {author} {\bibfnamefont {E.~W.}\ \bibnamefont
  {Brown}}, \bibinfo {author} {\bibfnamefont {J.~L.}\ \bibnamefont {{DuBois}}},
  \bibinfo {author} {\bibfnamefont {M.}~\bibnamefont {Holzmann}}, \ and\
  \bibinfo {author} {\bibfnamefont {D.~M.}\ \bibnamefont {Ceperley}},\ }\href
  {\doibase 10.1103/PhysRevB.88.081102} {\bibfield  {journal} {\bibinfo
  {journal} {Phys. Rev. B}\ }\textbf {\bibinfo {volume} {88}},\ \bibinfo
  {pages} {081102} (\bibinfo {year} {2013})}\BibitemShut {NoStop}%
\bibitem [{\citenamefont {Macke}(1950)}]{macke_uber_1950}%
  \BibitemOpen
  \bibfield  {author} {\bibinfo {author} {\bibfnamefont {V.~W.}\ \bibnamefont
  {Macke}},\ }\href@noop {} {\bibfield  {journal} {\bibinfo  {journal} {z.
  Naturforschg.}\ }\textbf {\bibinfo {volume} {5a}},\ \bibinfo {pages} {192}
  (\bibinfo {year} {1950})}\BibitemShut {NoStop}%
\bibitem [{\citenamefont {Gell-Mann}\ and\ \citenamefont
  {Brueckner}(1957)}]{gell-mann_correlation_1957}%
  \BibitemOpen
  \bibfield  {author} {\bibinfo {author} {\bibfnamefont {M.}~\bibnamefont
  {Gell-Mann}}\ and\ \bibinfo {author} {\bibfnamefont {K.~A.}\ \bibnamefont
  {Brueckner}},\ }\href {\doibase 10.1103/PhysRev.106.364} {\bibfield
  {journal} {\bibinfo  {journal} {Phys. Rev.}\ }\textbf {\bibinfo {volume}
  {106}},\ \bibinfo {pages} {364} (\bibinfo {year} {1957})}\BibitemShut
  {NoStop}%
\bibitem [{\citenamefont {Mattuck}(1992)}]{mattuck_guide_1992}%
  \BibitemOpen
  \bibfield  {author} {\bibinfo {author} {\bibfnamefont {R.~D.}\ \bibnamefont
  {Mattuck}},\ }\href@noop {} {\emph {\bibinfo {title} {A guide to Feynman
  diagrams in the many-body problem}}},\ \bibinfo {edition} {2nd}\ ed.,\ Dover
  books on physics and chemistry\ (\bibinfo  {publisher} {Dover Publications},\
  \bibinfo {address} {New York},\ \bibinfo {year} {1992})\BibitemShut {NoStop}%
\bibitem [{\citenamefont {Hirata}\ and\ \citenamefont
  {Ohnishi}(2012)}]{hirata_thermodynamic_2012}%
  \BibitemOpen
  \bibfield  {author} {\bibinfo {author} {\bibfnamefont {S.}~\bibnamefont
  {Hirata}}\ and\ \bibinfo {author} {\bibfnamefont {Y.-y.}\ \bibnamefont
  {Ohnishi}},\ }\href {\doibase 10.1039/C2CP23958B} {\bibfield  {journal}
  {\bibinfo  {journal} {Phys. Chem. Chem. Phys.}\ }\textbf {\bibinfo {volume}
  {14}},\ \bibinfo {pages} {7800} (\bibinfo {year} {2012})}\BibitemShut
  {NoStop}%
\bibitem [{\citenamefont {Bohm}\ and\ \citenamefont
  {Pines}(1953)}]{bohm_collective_1953}%
  \BibitemOpen
  \bibfield  {author} {\bibinfo {author} {\bibfnamefont {D.}~\bibnamefont
  {Bohm}}\ and\ \bibinfo {author} {\bibfnamefont {D.}~\bibnamefont {Pines}},\
  }\href {\doibase 10.1103/PhysRev.92.609} {\bibfield  {journal} {\bibinfo
  {journal} {Phys. Rev.}\ }\textbf {\bibinfo {volume} {92}},\ \bibinfo {pages}
  {609} (\bibinfo {year} {1953})}\BibitemShut {NoStop}%
\bibitem [{\citenamefont {Pines}\ and\ \citenamefont
  {Bohm}(1952)}]{pines_collective_1952}%
  \BibitemOpen
  \bibfield  {author} {\bibinfo {author} {\bibfnamefont {D.}~\bibnamefont
  {Pines}}\ and\ \bibinfo {author} {\bibfnamefont {D.}~\bibnamefont {Bohm}},\
  }\href {\doibase 10.1103/PhysRev.85.338} {\bibfield  {journal} {\bibinfo
  {journal} {Phys. Rev.}\ }\textbf {\bibinfo {volume} {85}},\ \bibinfo {pages}
  {338} (\bibinfo {year} {1952})}\BibitemShut {NoStop}%
\bibitem [{\citenamefont {Bohm}\ and\ \citenamefont
  {Pines}(1951)}]{bohm_collective_1951}%
  \BibitemOpen
  \bibfield  {author} {\bibinfo {author} {\bibfnamefont {D.}~\bibnamefont
  {Bohm}}\ and\ \bibinfo {author} {\bibfnamefont {D.}~\bibnamefont {Pines}},\
  }\href {\doibase 10.1103/PhysRev.82.625} {\bibfield  {journal} {\bibinfo
  {journal} {Phys. Rev.}\ }\textbf {\bibinfo {volume} {82}},\ \bibinfo {pages}
  {625} (\bibinfo {year} {1951})}\BibitemShut {NoStop}%
\bibitem [{\citenamefont {Freeman}(1977)}]{freeman_coupled-cluster_1977}%
  \BibitemOpen
  \bibfield  {author} {\bibinfo {author} {\bibfnamefont {D.~L.}\ \bibnamefont
  {Freeman}},\ }\href {\doibase 10.1103/PhysRevB.15.5512} {\bibfield  {journal}
  {\bibinfo  {journal} {Phys. Rev. B}\ }\textbf {\bibinfo {volume} {15}},\
  \bibinfo {pages} {5512} (\bibinfo {year} {1977})}\BibitemShut {NoStop}%
\bibitem [{\citenamefont {Bishop}\ and\ \citenamefont
  {L{\"u}hrmann}(1978)}]{bishop_electron_1978}%
  \BibitemOpen
  \bibfield  {author} {\bibinfo {author} {\bibfnamefont {R.~F.}\ \bibnamefont
  {Bishop}}\ and\ \bibinfo {author} {\bibfnamefont {K.~H.}\ \bibnamefont
  {L{\"u}hrmann}},\ }\href {\doibase 10.1103/PhysRevB.17.3757} {\bibfield
  {journal} {\bibinfo  {journal} {Phys. Rev. B}\ }\textbf {\bibinfo {volume}
  {17}},\ \bibinfo {pages} {3757} (\bibinfo {year} {1978})}\BibitemShut
  {NoStop}%
\bibitem [{\citenamefont {Bishop}\ and\ \citenamefont
  {L{\"u}hrmann}(1982)}]{bishop_electron_1982}%
  \BibitemOpen
  \bibfield  {author} {\bibinfo {author} {\bibfnamefont {R.~F.}\ \bibnamefont
  {Bishop}}\ and\ \bibinfo {author} {\bibfnamefont {K.~H.}\ \bibnamefont
  {L{\"u}hrmann}},\ }\href {\doibase 10.1103/PhysRevB.26.5523} {\bibfield
  {journal} {\bibinfo  {journal} {Phys. Rev. B}\ }\textbf {\bibinfo {volume}
  {26}},\ \bibinfo {pages} {5523} (\bibinfo {year} {1982})}\BibitemShut
  {NoStop}%
\bibitem [{\citenamefont {Bishop}(1991)}]{bishop_overview_1991}%
  \BibitemOpen
  \bibfield  {author} {\bibinfo {author} {\bibfnamefont {R.~F.}\ \bibnamefont
  {Bishop}},\ }\href {\doibase 10.1007/BF01119617} {\bibfield  {journal}
  {\bibinfo  {journal} {Theoretica chimica acta}\ }\textbf {\bibinfo {volume}
  {80}},\ \bibinfo {pages} {95} (\bibinfo {year} {1991})}\BibitemShut {NoStop}%
\bibitem [{\citenamefont {Bishop}\ \emph
  {et~al.}(1988{\natexlab{a}})\citenamefont {Bishop}, \citenamefont
  {Piechocki},\ and\ \citenamefont {Stevens}}]{bishop_pairing_1988}%
  \BibitemOpen
  \bibfield  {author} {\bibinfo {author} {\bibfnamefont {R.~F.}\ \bibnamefont
  {Bishop}}, \bibinfo {author} {\bibfnamefont {W.}~\bibnamefont {Piechocki}}, \
  and\ \bibinfo {author} {\bibfnamefont {G.~A.}\ \bibnamefont {Stevens}},\
  }\href {\doibase 10.1007/BF01075347} {\bibfield  {journal} {\bibinfo
  {journal} {Few-Body Systems}\ }\textbf {\bibinfo {volume} {4}},\ \bibinfo
  {pages} {161} (\bibinfo {year} {1988}{\natexlab{a}})}\BibitemShut {NoStop}%
\bibitem [{\citenamefont {Bishop}\ \emph
  {et~al.}(1988{\natexlab{b}})\citenamefont {Bishop}, \citenamefont
  {Piechocki},\ and\ \citenamefont {Stevens}}]{bishop_pairing_1988-1}%
  \BibitemOpen
  \bibfield  {author} {\bibinfo {author} {\bibfnamefont {R.~F.}\ \bibnamefont
  {Bishop}}, \bibinfo {author} {\bibfnamefont {W.}~\bibnamefont {Piechocki}}, \
  and\ \bibinfo {author} {\bibfnamefont {G.~A.}\ \bibnamefont {Stevens}},\
  }\href {\doibase 10.1007/BF01076993} {\bibfield  {journal} {\bibinfo
  {journal} {Few-Body Systems}\ }\textbf {\bibinfo {volume} {4}},\ \bibinfo
  {pages} {179} (\bibinfo {year} {1988}{\natexlab{b}})}\BibitemShut {NoStop}%
\bibitem [{\citenamefont {Furche}(2001)}]{furche_molecular_2001}%
  \BibitemOpen
  \bibfield  {author} {\bibinfo {author} {\bibfnamefont {F.}~\bibnamefont
  {Furche}},\ }\href {\doibase 10.1103/PhysRevB.64.195120} {\bibfield
  {journal} {\bibinfo  {journal} {Phys. Rev. B}\ }\textbf {\bibinfo {volume}
  {64}},\ \bibinfo {pages} {195120} (\bibinfo {year} {2001})}\BibitemShut
  {NoStop}%
\bibitem [{\citenamefont {He{\ss}elmann}\ and\ \citenamefont
  {G{\"o}rling}(2011)}]{heselmann_random-phase_2011}%
  \BibitemOpen
  \bibfield  {author} {\bibinfo {author} {\bibfnamefont {A.}~\bibnamefont
  {He{\ss}elmann}}\ and\ \bibinfo {author} {\bibfnamefont {A.}~\bibnamefont
  {G{\"o}rling}},\ }\href {\doibase 10.1080/00268976.2011.614282} {\bibfield
  {journal} {\bibinfo  {journal} {Mol. Phys.}\ }\textbf {\bibinfo {volume}
  {109}},\ \bibinfo {pages} {2473} (\bibinfo {year} {2011})}\BibitemShut
  {NoStop}%
\bibitem [{\citenamefont
  {Furche}(2008{\natexlab{a}})}]{furche_developing_2008}%
  \BibitemOpen
  \bibfield  {author} {\bibinfo {author} {\bibfnamefont {F.}~\bibnamefont
  {Furche}},\ }\href {\doibase 10.1063/1.2977789} {\bibfield  {journal}
  {\bibinfo  {journal} {J. Chem. Phys.}\ }\textbf {\bibinfo {volume} {129}},\
  \bibinfo {pages} {114105} (\bibinfo {year} {2008}{\natexlab{a}})}\BibitemShut
  {NoStop}%
\bibitem [{\citenamefont
  {Furche}(2008{\natexlab{b}})}]{furche_developing_2008-1}%
  \BibitemOpen
  \bibfield  {author} {\bibinfo {author} {\bibfnamefont {F.}~\bibnamefont
  {Furche}},\ }\href {\doibase 10.1063/1.2977789} {\bibfield  {journal}
  {\bibinfo  {journal} {J. Chem. Phys.}\ }\textbf {\bibinfo {volume} {129}},\
  \bibinfo {pages} {114105} (\bibinfo {year} {2008}{\natexlab{b}})}\BibitemShut
  {NoStop}%
\bibitem [{\citenamefont {Eshuis}\ \emph {et~al.}(2010)\citenamefont {Eshuis},
  \citenamefont {Yarkony},\ and\ \citenamefont {Furche}}]{eshuis_fast_2010}%
  \BibitemOpen
  \bibfield  {author} {\bibinfo {author} {\bibfnamefont {H.}~\bibnamefont
  {Eshuis}}, \bibinfo {author} {\bibfnamefont {J.}~\bibnamefont {Yarkony}}, \
  and\ \bibinfo {author} {\bibfnamefont {F.}~\bibnamefont {Furche}},\ }\href
  {\doibase 10.1063/1.3442749} {\bibfield  {journal} {\bibinfo  {journal} {J.
  Chem. Phys.}\ }\textbf {\bibinfo {volume} {132}},\ \bibinfo {pages} {234114}
  (\bibinfo {year} {2010})}\BibitemShut {NoStop}%
\bibitem [{\citenamefont {Eshuis}\ and\ \citenamefont
  {Furche}(2012)}]{eshuis_basis_2012}%
  \BibitemOpen
  \bibfield  {author} {\bibinfo {author} {\bibfnamefont {H.}~\bibnamefont
  {Eshuis}}\ and\ \bibinfo {author} {\bibfnamefont {F.}~\bibnamefont
  {Furche}},\ }\href {\doibase 10.1063/1.3687005} {\bibfield  {journal}
  {\bibinfo  {journal} {J. Chem. Phys.}\ }\textbf {\bibinfo {volume} {136}},\
  \bibinfo {pages} {084105} (\bibinfo {year} {2012})}\BibitemShut {NoStop}%
\bibitem [{\citenamefont {Paier}\ \emph {et~al.}(2010)\citenamefont {Paier},
  \citenamefont {Janesko}, \citenamefont {Henderson}, \citenamefont {Scuseria},
  \citenamefont {Gr{\"u}neis},\ and\ \citenamefont
  {Kresse}}]{paier_hybrid_2010}%
  \BibitemOpen
  \bibfield  {author} {\bibinfo {author} {\bibfnamefont {J.}~\bibnamefont
  {Paier}}, \bibinfo {author} {\bibfnamefont {B.~G.}\ \bibnamefont {Janesko}},
  \bibinfo {author} {\bibfnamefont {T.~M.}\ \bibnamefont {Henderson}}, \bibinfo
  {author} {\bibfnamefont {G.~E.}\ \bibnamefont {Scuseria}}, \bibinfo {author}
  {\bibfnamefont {A.}~\bibnamefont {Gr{\"u}neis}}, \ and\ \bibinfo {author}
  {\bibfnamefont {G.}~\bibnamefont {Kresse}},\ }\href {\doibase
  doi:10.1063/1.3317437} {\bibfield  {journal} {\bibinfo  {journal} {J. Chem.
  Phys.}\ }\textbf {\bibinfo {volume} {132}},\ \bibinfo {pages} {094103}
  (\bibinfo {year} {2010})}\BibitemShut {NoStop}%
\bibitem [{\citenamefont {Janesko}\ \emph
  {et~al.}(2009{\natexlab{a}})\citenamefont {Janesko}, \citenamefont
  {Henderson},\ and\ \citenamefont
  {Scuseria}}]{janesko_long-range-corrected_2009}%
  \BibitemOpen
  \bibfield  {author} {\bibinfo {author} {\bibfnamefont {B.~G.}\ \bibnamefont
  {Janesko}}, \bibinfo {author} {\bibfnamefont {T.~M.}\ \bibnamefont
  {Henderson}}, \ and\ \bibinfo {author} {\bibfnamefont {G.~E.}\ \bibnamefont
  {Scuseria}},\ }\href {\doibase doi:10.1063/1.3090814} {\bibfield  {journal}
  {\bibinfo  {journal} {J. Chem. Phys.}\ }\textbf {\bibinfo {volume} {130}},\
  \bibinfo {pages} {081105} (\bibinfo {year} {2009}{\natexlab{a}})}\BibitemShut
  {NoStop}%
\bibitem [{\citenamefont {Janesko}\ \emph
  {et~al.}(2009{\natexlab{b}})\citenamefont {Janesko}, \citenamefont
  {Henderson},\ and\ \citenamefont
  {Scuseria}}]{janesko_long-range-corrected_2009-1}%
  \BibitemOpen
  \bibfield  {author} {\bibinfo {author} {\bibfnamefont {B.~G.}\ \bibnamefont
  {Janesko}}, \bibinfo {author} {\bibfnamefont {T.~M.}\ \bibnamefont
  {Henderson}}, \ and\ \bibinfo {author} {\bibfnamefont {G.~E.}\ \bibnamefont
  {Scuseria}},\ }\href {\doibase doi:10.1063/1.3176514} {\bibfield  {journal}
  {\bibinfo  {journal} {J. Chem. Phys.}\ }\textbf {\bibinfo {volume} {131}},\
  \bibinfo {pages} {034110} (\bibinfo {year} {2009}{\natexlab{b}})}\BibitemShut
  {NoStop}%
\bibitem [{\citenamefont {Henderson}\ and\ \citenamefont
  {Scuseria}(2010)}]{henderson_connection_2010}%
  \BibitemOpen
  \bibfield  {author} {\bibinfo {author} {\bibfnamefont {T.~M.}\ \bibnamefont
  {Henderson}}\ and\ \bibinfo {author} {\bibfnamefont {G.~E.}\ \bibnamefont
  {Scuseria}},\ }\href {\doibase 10.1080/00268976.2010.507227} {\bibfield
  {journal} {\bibinfo  {journal} {Mol. Phys.}\ }\textbf {\bibinfo {volume}
  {108}},\ \bibinfo {pages} {2511} (\bibinfo {year} {2010})}\BibitemShut
  {NoStop}%
\bibitem [{\citenamefont {Irelan}\ \emph {et~al.}(2011)\citenamefont {Irelan},
  \citenamefont {Henderson},\ and\ \citenamefont
  {Scuseria}}]{irelan_long-range-corrected_2011}%
  \BibitemOpen
  \bibfield  {author} {\bibinfo {author} {\bibfnamefont {R.~M.}\ \bibnamefont
  {Irelan}}, \bibinfo {author} {\bibfnamefont {T.~M.}\ \bibnamefont
  {Henderson}}, \ and\ \bibinfo {author} {\bibfnamefont {G.~E.}\ \bibnamefont
  {Scuseria}},\ }\href {\doibase doi:10.1063/1.3630951} {\bibfield  {journal}
  {\bibinfo  {journal} {J. Chem. Phys.}\ }\textbf {\bibinfo {volume} {135}},\
  \bibinfo {pages} {094105} (\bibinfo {year} {2011})}\BibitemShut {NoStop}%
\bibitem [{\citenamefont {Janesko}\ and\ \citenamefont
  {Scuseria}(2009)}]{janesko_role_2009}%
  \BibitemOpen
  \bibfield  {author} {\bibinfo {author} {\bibfnamefont {B.~G.}\ \bibnamefont
  {Janesko}}\ and\ \bibinfo {author} {\bibfnamefont {G.~E.}\ \bibnamefont
  {Scuseria}},\ }\href {\doibase 10.1063/1.3250834} {\bibfield  {journal}
  {\bibinfo  {journal} {J. Chem. Phys.}\ }\textbf {\bibinfo {volume} {131}},\
  \bibinfo {pages} {154106} (\bibinfo {year} {2009})}\BibitemShut {NoStop}%
\bibitem [{\citenamefont {Eshuis}\ \emph {et~al.}(2012)\citenamefont {Eshuis},
  \citenamefont {Bates},\ and\ \citenamefont {Furche}}]{eshuis_electron_2012}%
  \BibitemOpen
  \bibfield  {author} {\bibinfo {author} {\bibfnamefont {H.}~\bibnamefont
  {Eshuis}}, \bibinfo {author} {\bibfnamefont {J.~E.}\ \bibnamefont {Bates}}, \
  and\ \bibinfo {author} {\bibfnamefont {F.}~\bibnamefont {Furche}},\ }\href
  {\doibase 10.1007/s00214-011-1084-8} {\bibfield  {journal} {\bibinfo
  {journal} {Theor. Chem. Acc.}\ }\textbf {\bibinfo {volume} {131}},\ \bibinfo
  {pages} {1} (\bibinfo {year} {2012})}\BibitemShut {NoStop}%
\bibitem [{\citenamefont {Ren}\ \emph {et~al.}(2012)\citenamefont {Ren},
  \citenamefont {Rinke}, \citenamefont {Joas},\ and\ \citenamefont
  {Scheffler}}]{ren_random-phase_2012}%
  \BibitemOpen
  \bibfield  {author} {\bibinfo {author} {\bibfnamefont {X.}~\bibnamefont
  {Ren}}, \bibinfo {author} {\bibfnamefont {P.}~\bibnamefont {Rinke}}, \bibinfo
  {author} {\bibfnamefont {C.}~\bibnamefont {Joas}}, \ and\ \bibinfo {author}
  {\bibfnamefont {M.}~\bibnamefont {Scheffler}},\ }\href {\doibase
  10.1007/s10853-012-6570-4} {\bibfield  {journal} {\bibinfo  {journal}
  {Journal of Materials Science}\ }\textbf {\bibinfo {volume} {47}},\ \bibinfo
  {pages} {7447} (\bibinfo {year} {2012})}\BibitemShut {NoStop}%
\bibitem [{\citenamefont {Paier}\ \emph {et~al.}(2012)\citenamefont {Paier},
  \citenamefont {Ren}, \citenamefont {Rinke}, \citenamefont {Scuseria},
  \citenamefont {Gr{\"u}neis}, \citenamefont {Kresse},\ and\ \citenamefont
  {Scheffler}}]{paier_assessment_2012}%
  \BibitemOpen
  \bibfield  {author} {\bibinfo {author} {\bibfnamefont {J.}~\bibnamefont
  {Paier}}, \bibinfo {author} {\bibfnamefont {X.}~\bibnamefont {Ren}}, \bibinfo
  {author} {\bibfnamefont {P.}~\bibnamefont {Rinke}}, \bibinfo {author}
  {\bibfnamefont {G.~E.}\ \bibnamefont {Scuseria}}, \bibinfo {author}
  {\bibfnamefont {A.}~\bibnamefont {Gr{\"u}neis}}, \bibinfo {author}
  {\bibfnamefont {G.}~\bibnamefont {Kresse}}, \ and\ \bibinfo {author}
  {\bibfnamefont {M.}~\bibnamefont {Scheffler}},\ }\href {\doibase
  10.1088/1367-2630/14/4/043002} {\bibfield  {journal} {\bibinfo  {journal}
  {New J. Phys.}\ }\textbf {\bibinfo {volume} {14}},\ \bibinfo {pages} {043002}
  (\bibinfo {year} {2012})}\BibitemShut {NoStop}%
\bibitem [{\citenamefont {Dobson}\ \emph {et~al.}(2006)\citenamefont {Dobson},
  \citenamefont {White},\ and\ \citenamefont
  {Rubio}}]{dobson_asymptotics_2006}%
  \BibitemOpen
  \bibfield  {author} {\bibinfo {author} {\bibfnamefont {J.~F.}\ \bibnamefont
  {Dobson}}, \bibinfo {author} {\bibfnamefont {A.}~\bibnamefont {White}}, \
  and\ \bibinfo {author} {\bibfnamefont {A.}~\bibnamefont {Rubio}},\ }\href
  {\doibase 10.1103/PhysRevLett.96.073201} {\bibfield  {journal} {\bibinfo
  {journal} {Phys. Rev. Lett.}\ }\textbf {\bibinfo {volume} {96}},\ \bibinfo
  {pages} {073201} (\bibinfo {year} {2006})}\BibitemShut {NoStop}%
\bibitem [{\citenamefont {Harl}\ and\ \citenamefont
  {Kresse}(2009)}]{harl_accurate_2009}%
  \BibitemOpen
  \bibfield  {author} {\bibinfo {author} {\bibfnamefont {J.}~\bibnamefont
  {Harl}}\ and\ \bibinfo {author} {\bibfnamefont {G.}~\bibnamefont {Kresse}},\
  }\href {\doibase 10.1103/PhysRevLett.103.056401} {\bibfield  {journal}
  {\bibinfo  {journal} {Phys. Rev. Lett.}\ }\textbf {\bibinfo {volume} {103}},\
  \bibinfo {pages} {056401} (\bibinfo {year} {2009})}\BibitemShut {NoStop}%
\bibitem [{\citenamefont {Leb{\`e}gue}\ \emph {et~al.}(2010)\citenamefont
  {Leb{\`e}gue}, \citenamefont {Harl}, \citenamefont {Gould}, \citenamefont
  {{\'A}ngy{\'a}n}, \citenamefont {Kresse},\ and\ \citenamefont
  {Dobson}}]{lebegue_cohesive_2010}%
  \BibitemOpen
  \bibfield  {author} {\bibinfo {author} {\bibfnamefont {S.}~\bibnamefont
  {Leb{\`e}gue}}, \bibinfo {author} {\bibfnamefont {J.}~\bibnamefont {Harl}},
  \bibinfo {author} {\bibfnamefont {T.}~\bibnamefont {Gould}}, \bibinfo
  {author} {\bibfnamefont {J.~G.}\ \bibnamefont {{\'A}ngy{\'a}n}}, \bibinfo
  {author} {\bibfnamefont {G.}~\bibnamefont {Kresse}}, \ and\ \bibinfo {author}
  {\bibfnamefont {J.~F.}\ \bibnamefont {Dobson}},\ }\href {\doibase
  10.1103/PhysRevLett.105.196401} {\bibfield  {journal} {\bibinfo  {journal}
  {Phys. Rev. Lett.}\ }\textbf {\bibinfo {volume} {105}},\ \bibinfo {pages}
  {196401} (\bibinfo {year} {2010})}\BibitemShut {NoStop}%
\bibitem [{\citenamefont {Toulouse}\ \emph {et~al.}(2011)\citenamefont
  {Toulouse}, \citenamefont {Zhu}, \citenamefont {Savin}, \citenamefont
  {Jansen},\ and\ \citenamefont {{\'A}ngy{\'a}n}}]{toulouse_closed-shell_2011}%
  \BibitemOpen
  \bibfield  {author} {\bibinfo {author} {\bibfnamefont {J.}~\bibnamefont
  {Toulouse}}, \bibinfo {author} {\bibfnamefont {W.}~\bibnamefont {Zhu}},
  \bibinfo {author} {\bibfnamefont {A.}~\bibnamefont {Savin}}, \bibinfo
  {author} {\bibfnamefont {G.}~\bibnamefont {Jansen}}, \ and\ \bibinfo {author}
  {\bibfnamefont {J.~G.}\ \bibnamefont {{\'A}ngy{\'a}n}},\ }\href {\doibase
  doi:10.1063/1.3626551} {\bibfield  {journal} {\bibinfo  {journal} {J. Chem.
  Phys.}\ }\textbf {\bibinfo {volume} {135}},\ \bibinfo {pages} {084119}
  (\bibinfo {year} {2011})}\BibitemShut {NoStop}%
\bibitem [{\citenamefont {{\'A}ngy{\'a}n}\ \emph {et~al.}(2011)\citenamefont
  {{\'A}ngy{\'a}n}, \citenamefont {Liu}, \citenamefont {Toulouse},\ and\
  \citenamefont {Jansen}}]{angyan_correlation_2011}%
  \BibitemOpen
  \bibfield  {author} {\bibinfo {author} {\bibfnamefont {J.~G.}\ \bibnamefont
  {{\'A}ngy{\'a}n}}, \bibinfo {author} {\bibfnamefont {R.-F.}\ \bibnamefont
  {Liu}}, \bibinfo {author} {\bibfnamefont {J.}~\bibnamefont {Toulouse}}, \
  and\ \bibinfo {author} {\bibfnamefont {G.}~\bibnamefont {Jansen}},\ }\href
  {\doibase 10.1021/ct200501r} {\bibfield  {journal} {\bibinfo  {journal} {J.
  Chem. Theory Comput.}\ }\textbf {\bibinfo {volume} {7}},\ \bibinfo {pages}
  {3116} (\bibinfo {year} {2011})}\BibitemShut {NoStop}%
\bibitem [{\citenamefont {Toulouse}\ \emph {et~al.}(2009)\citenamefont
  {Toulouse}, \citenamefont {Gerber}, \citenamefont {Jansen}, \citenamefont
  {Savin},\ and\ \citenamefont
  {{\'A}ngy{\'a}n}}]{toulouse_adiabatic-connection_2009}%
  \BibitemOpen
  \bibfield  {author} {\bibinfo {author} {\bibfnamefont {J.}~\bibnamefont
  {Toulouse}}, \bibinfo {author} {\bibfnamefont {I.~C.}\ \bibnamefont
  {Gerber}}, \bibinfo {author} {\bibfnamefont {G.}~\bibnamefont {Jansen}},
  \bibinfo {author} {\bibfnamefont {A.}~\bibnamefont {Savin}}, \ and\ \bibinfo
  {author} {\bibfnamefont {J.~G.}\ \bibnamefont {{\'A}ngy{\'a}n}},\ }\href
  {\doibase 10.1103/PhysRevLett.102.096404} {\bibfield  {journal} {\bibinfo
  {journal} {Phys. Rev. Lett.}\ }\textbf {\bibinfo {volume} {102}},\ \bibinfo
  {pages} {096404} (\bibinfo {year} {2009})}\BibitemShut {NoStop}%
\bibitem [{\citenamefont {Calmels}\ and\ \citenamefont
  {Gold}(1998)}]{calmels_pair-correlation_1998}%
  \BibitemOpen
  \bibfield  {author} {\bibinfo {author} {\bibfnamefont {L.}~\bibnamefont
  {Calmels}}\ and\ \bibinfo {author} {\bibfnamefont {A.}~\bibnamefont {Gold}},\
  }\href {\doibase 10.1103/PhysRevB.57.1436} {\bibfield  {journal} {\bibinfo
  {journal} {Phys. Rev. B}\ }\textbf {\bibinfo {volume} {57}},\ \bibinfo
  {pages} {1436} (\bibinfo {year} {1998})}\BibitemShut {NoStop}%
\bibitem [{\citenamefont {Nagano}\ \emph {et~al.}(1984)\citenamefont {Nagano},
  \citenamefont {Singwi},\ and\ \citenamefont
  {Ohnishi}}]{nagano_correlations_1984}%
  \BibitemOpen
  \bibfield  {author} {\bibinfo {author} {\bibfnamefont {S.}~\bibnamefont
  {Nagano}}, \bibinfo {author} {\bibfnamefont {K.~S.}\ \bibnamefont {Singwi}},
  \ and\ \bibinfo {author} {\bibfnamefont {S.}~\bibnamefont {Ohnishi}},\ }\href
  {\doibase 10.1103/PhysRevB.29.1209} {\bibfield  {journal} {\bibinfo
  {journal} {Phys. Rev. B}\ }\textbf {\bibinfo {volume} {29}},\ \bibinfo
  {pages} {1209} (\bibinfo {year} {1984})}\BibitemShut {NoStop}%
\bibitem [{\citenamefont {Yasuhara}(1972)}]{yasuhara_short-range_1972}%
  \BibitemOpen
  \bibfield  {author} {\bibinfo {author} {\bibfnamefont {H.}~\bibnamefont
  {Yasuhara}},\ }\href {\doibase 10.1016/0038-1098(72)90504-2} {\bibfield
  {journal} {\bibinfo  {journal} {Solid State Commun.}\ }\textbf {\bibinfo
  {volume} {11}},\ \bibinfo {pages} {1481} (\bibinfo {year}
  {1972})}\BibitemShut {NoStop}%
\bibitem [{\citenamefont
  {Yasuhara}(1974{\natexlab{a}})}]{yasuhara_electron_1974}%
  \BibitemOpen
  \bibfield  {author} {\bibinfo {author} {\bibfnamefont {H.}~\bibnamefont
  {Yasuhara}},\ }\href {\doibase 10.1143/JPSJ.36.361} {\bibfield  {journal}
  {\bibinfo  {journal} {J. Phys. Soc. Jpn.}\ }\textbf {\bibinfo {volume}
  {36}},\ \bibinfo {pages} {361} (\bibinfo {year}
  {1974}{\natexlab{a}})}\BibitemShut {NoStop}%
\bibitem [{\citenamefont
  {Yasuhara}(1974{\natexlab{b}})}]{yasuhara_erratum:_1974}%
  \BibitemOpen
  \bibfield  {author} {\bibinfo {author} {\bibfnamefont {H.}~\bibnamefont
  {Yasuhara}},\ }\href {\doibase 10.1143/JPSJ.37.579A} {\bibfield  {journal}
  {\bibinfo  {journal} {J. Phys. Soc. Jpn.}\ }\textbf {\bibinfo {volume}
  {37}},\ \bibinfo {pages} {579A} (\bibinfo {year}
  {1974}{\natexlab{b}})}\BibitemShut {NoStop}%
\bibitem [{\citenamefont {Lowy}\ and\ \citenamefont
  {Brown}(1975)}]{lowy_electron_1975}%
  \BibitemOpen
  \bibfield  {author} {\bibinfo {author} {\bibfnamefont {D.~N.}\ \bibnamefont
  {Lowy}}\ and\ \bibinfo {author} {\bibfnamefont {G.~E.}\ \bibnamefont
  {Brown}},\ }\href {\doibase 10.1103/PhysRevB.12.2138} {\bibfield  {journal}
  {\bibinfo  {journal} {Phys. Rev. B}\ }\textbf {\bibinfo {volume} {12}},\
  \bibinfo {pages} {2138} (\bibinfo {year} {1975})}\BibitemShut {NoStop}%
\bibitem [{\citenamefont {Bedell}\ and\ \citenamefont
  {Brown}(1978)}]{bedell_short-range_1978}%
  \BibitemOpen
  \bibfield  {author} {\bibinfo {author} {\bibfnamefont {K.}~\bibnamefont
  {Bedell}}\ and\ \bibinfo {author} {\bibfnamefont {G.~E.}\ \bibnamefont
  {Brown}},\ }\href {\doibase 10.1103/PhysRevB.17.4512} {\bibfield  {journal}
  {\bibinfo  {journal} {Phys. Rev. B}\ }\textbf {\bibinfo {volume} {17}},\
  \bibinfo {pages} {4512} (\bibinfo {year} {1978})}\BibitemShut {NoStop}%
\bibitem [{\citenamefont {Qian}(2006)}]{qian_-top_2006}%
  \BibitemOpen
  \bibfield  {author} {\bibinfo {author} {\bibfnamefont {Z.}~\bibnamefont
  {Qian}},\ }\href {\doibase 10.1103/PhysRevB.73.035106} {\bibfield  {journal}
  {\bibinfo  {journal} {Phys. Rev. B}\ }\textbf {\bibinfo {volume} {73}},\
  \bibinfo {pages} {035106} (\bibinfo {year} {2006})}\BibitemShut {NoStop}%
\bibitem [{\citenamefont {Drummond}\ and\ \citenamefont
  {Needs}(2009)}]{drummond_quantum_2009}%
  \BibitemOpen
  \bibfield  {author} {\bibinfo {author} {\bibfnamefont {N.~D.}\ \bibnamefont
  {Drummond}}\ and\ \bibinfo {author} {\bibfnamefont {R.~J.}\ \bibnamefont
  {Needs}},\ }\href {\doibase 10.1103/PhysRevB.79.085414} {\bibfield  {journal}
  {\bibinfo  {journal} {Phys. Rev. B}\ }\textbf {\bibinfo {volume} {79}},\
  \bibinfo {pages} {085414} (\bibinfo {year} {2009})}\BibitemShut {NoStop}%
\bibitem [{\citenamefont {Awa}\ and\ \citenamefont
  {Asahi}(1980)}]{awa_electron_1980}%
  \BibitemOpen
  \bibfield  {author} {\bibinfo {author} {\bibfnamefont {K.}~\bibnamefont
  {Awa}}\ and\ \bibinfo {author} {\bibfnamefont {T.}~\bibnamefont {Asahi}},\
  }\href {\doibase 10.1143/JPSJ.48.757} {\bibfield  {journal} {\bibinfo
  {journal} {J. Phys. Soc. Jpn.}\ }\textbf {\bibinfo {volume} {48}},\ \bibinfo
  {pages} {757} (\bibinfo {year} {1980})}\BibitemShut {NoStop}%
\bibitem [{\citenamefont {Yasuhara}\ \emph {et~al.}(1988)\citenamefont
  {Yasuhara}, \citenamefont {Suehiro},\ and\ \citenamefont
  {Ousaka}}]{yasuhara_new_1988}%
  \BibitemOpen
  \bibfield  {author} {\bibinfo {author} {\bibfnamefont {H.}~\bibnamefont
  {Yasuhara}}, \bibinfo {author} {\bibfnamefont {H.}~\bibnamefont {Suehiro}}, \
  and\ \bibinfo {author} {\bibfnamefont {Y.}~\bibnamefont {Ousaka}},\ }\href
  {\doibase 10.1088/0022-3719/21/22/019} {\bibfield  {journal} {\bibinfo
  {journal} {Journal of Physics C: Solid State Physics}\ }\textbf {\bibinfo
  {volume} {21}},\ \bibinfo {pages} {4045} (\bibinfo {year}
  {1988})}\BibitemShut {NoStop}%
\bibitem [{\citenamefont {Hagen}\ \emph {et~al.}(2013)\citenamefont {Hagen},
  \citenamefont {Papenbrock}, \citenamefont {Ekstr{\"o}m}, \citenamefont
  {Wendt}, \citenamefont {Baardsen}, \citenamefont {Gandolfi}, \citenamefont
  {Hjorth-Jensen},\ and\ \citenamefont
  {Horowitz}}]{hagen_coupled-cluster_2013}%
  \BibitemOpen
  \bibfield  {author} {\bibinfo {author} {\bibfnamefont {G.}~\bibnamefont
  {Hagen}}, \bibinfo {author} {\bibfnamefont {T.}~\bibnamefont {Papenbrock}},
  \bibinfo {author} {\bibfnamefont {A.}~\bibnamefont {Ekstr{\"o}m}}, \bibinfo
  {author} {\bibfnamefont {K.~A.}\ \bibnamefont {Wendt}}, \bibinfo {author}
  {\bibfnamefont {G.}~\bibnamefont {Baardsen}}, \bibinfo {author}
  {\bibfnamefont {S.}~\bibnamefont {Gandolfi}}, \bibinfo {author}
  {\bibfnamefont {M.}~\bibnamefont {Hjorth-Jensen}}, \ and\ \bibinfo {author}
  {\bibfnamefont {C.~J.}\ \bibnamefont {Horowitz}},\ }\href
  {http://arxiv.org/abs/1311.2925} {\emph {\bibinfo {title} {Coupled-cluster
  calculations of nucleonic matter}}},\ \bibinfo {type} {{arXiv} e-print}\
  \bibinfo {number} {1311.2925}\ (\bibinfo {year} {2013})\BibitemShut {NoStop}%
\bibitem [{\citenamefont {Paier}\ \emph {et~al.}(2009)\citenamefont {Paier},
  \citenamefont {Diaconu}, \citenamefont {Scuseria}, \citenamefont {Guidon},
  \citenamefont {{VandeVondele}},\ and\ \citenamefont
  {Hutter}}]{paier_accurate_2009}%
  \BibitemOpen
  \bibfield  {author} {\bibinfo {author} {\bibfnamefont {J.}~\bibnamefont
  {Paier}}, \bibinfo {author} {\bibfnamefont {C.~V.}\ \bibnamefont {Diaconu}},
  \bibinfo {author} {\bibfnamefont {G.~E.}\ \bibnamefont {Scuseria}}, \bibinfo
  {author} {\bibfnamefont {M.}~\bibnamefont {Guidon}}, \bibinfo {author}
  {\bibfnamefont {J.}~\bibnamefont {{VandeVondele}}}, \ and\ \bibinfo {author}
  {\bibfnamefont {J.}~\bibnamefont {Hutter}},\ }\href {\doibase
  10.1103/PhysRevB.80.174114} {\bibfield  {journal} {\bibinfo  {journal} {Phys.
  Rev. B}\ }\textbf {\bibinfo {volume} {80}},\ \bibinfo {pages} {174114}
  (\bibinfo {year} {2009})}\BibitemShut {NoStop}%
\bibitem [{\citenamefont {Dovesi}\ \emph {et~al.}(1984)\citenamefont {Dovesi},
  \citenamefont {Pisani}, \citenamefont {Roetti},\ and\ \citenamefont
  {Saunders}}]{dovesi_hartreefock_1984}%
  \BibitemOpen
  \bibfield  {author} {\bibinfo {author} {\bibfnamefont {R.}~\bibnamefont
  {Dovesi}}, \bibinfo {author} {\bibfnamefont {C.}~\bibnamefont {Pisani}},
  \bibinfo {author} {\bibfnamefont {C.}~\bibnamefont {Roetti}}, \ and\ \bibinfo
  {author} {\bibfnamefont {V.~R.}\ \bibnamefont {Saunders}},\ }\href {\doibase
  10.1063/1.447957} {\bibfield  {journal} {\bibinfo  {journal} {J. Chem.
  Phys.}\ }\textbf {\bibinfo {volume} {81}},\ \bibinfo {pages} {2839} (\bibinfo
  {year} {1984})}\BibitemShut {NoStop}%
\bibitem [{\citenamefont {Gillan}\ \emph {et~al.}(2008)\citenamefont {Gillan},
  \citenamefont {Alf{\`e}}, \citenamefont {de~Gironcoli},\ and\ \citenamefont
  {Manby}}]{gillan_high-precision_2008}%
  \BibitemOpen
  \bibfield  {author} {\bibinfo {author} {\bibfnamefont {M.~J.}\ \bibnamefont
  {Gillan}}, \bibinfo {author} {\bibfnamefont {D.}~\bibnamefont {Alf{\`e}}},
  \bibinfo {author} {\bibfnamefont {S.}~\bibnamefont {de~Gironcoli}}, \ and\
  \bibinfo {author} {\bibfnamefont {F.~R.}\ \bibnamefont {Manby}},\ }\href
  {\doibase 10.1002/jcc.21033} {\bibfield  {journal} {\bibinfo  {journal}
  {Journal of Computational Chemistry}\ }\textbf {\bibinfo {volume} {29}},\
  \bibinfo {pages} {2098{\textendash}2106} (\bibinfo {year}
  {2008})}\BibitemShut {NoStop}%
\bibitem [{\citenamefont {Pisani}\ \emph {et~al.}(2008)\citenamefont {Pisani},
  \citenamefont {Maschio}, \citenamefont {Casassa}, \citenamefont {Halo},
  \citenamefont {Sch{\"u}tz},\ and\ \citenamefont
  {Usvyat}}]{pisani_periodic_2008}%
  \BibitemOpen
  \bibfield  {author} {\bibinfo {author} {\bibfnamefont {C.}~\bibnamefont
  {Pisani}}, \bibinfo {author} {\bibfnamefont {L.}~\bibnamefont {Maschio}},
  \bibinfo {author} {\bibfnamefont {S.}~\bibnamefont {Casassa}}, \bibinfo
  {author} {\bibfnamefont {M.}~\bibnamefont {Halo}}, \bibinfo {author}
  {\bibfnamefont {M.}~\bibnamefont {Sch{\"u}tz}}, \ and\ \bibinfo {author}
  {\bibfnamefont {D.}~\bibnamefont {Usvyat}},\ }\href {\doibase
  10.1002/jcc.20975} {\bibfield  {journal} {\bibinfo  {journal} {Journal of
  Computational Chemistry}\ }\textbf {\bibinfo {volume} {29}},\ \bibinfo
  {pages} {2113{\textendash}2124} (\bibinfo {year} {2008})}\BibitemShut
  {NoStop}%
\bibitem [{\citenamefont {Schwerdtfeger}\ \emph {et~al.}(2010)\citenamefont
  {Schwerdtfeger}, \citenamefont {Assadollahzadeh},\ and\ \citenamefont
  {Hermann}}]{schwerdtfeger_convergence_2010}%
  \BibitemOpen
  \bibfield  {author} {\bibinfo {author} {\bibfnamefont {P.}~\bibnamefont
  {Schwerdtfeger}}, \bibinfo {author} {\bibfnamefont {B.}~\bibnamefont
  {Assadollahzadeh}}, \ and\ \bibinfo {author} {\bibfnamefont {A.}~\bibnamefont
  {Hermann}},\ }\href {\doibase 10.1103/PhysRevB.82.205111} {\bibfield
  {journal} {\bibinfo  {journal} {Phys. Rev. B}\ }\textbf {\bibinfo {volume}
  {82}},\ \bibinfo {pages} {205111} (\bibinfo {year} {2010})}\BibitemShut
  {NoStop}%
\bibitem [{\citenamefont {Maschio}\ \emph {et~al.}(2007)\citenamefont
  {Maschio}, \citenamefont {Usvyat}, \citenamefont {Manby}, \citenamefont
  {Casassa}, \citenamefont {Pisani},\ and\ \citenamefont
  {Sch{\"u}tz}}]{maschio_fast_2007}%
  \BibitemOpen
  \bibfield  {author} {\bibinfo {author} {\bibfnamefont {L.}~\bibnamefont
  {Maschio}}, \bibinfo {author} {\bibfnamefont {D.}~\bibnamefont {Usvyat}},
  \bibinfo {author} {\bibfnamefont {F.~R.}\ \bibnamefont {Manby}}, \bibinfo
  {author} {\bibfnamefont {S.}~\bibnamefont {Casassa}}, \bibinfo {author}
  {\bibfnamefont {C.}~\bibnamefont {Pisani}}, \ and\ \bibinfo {author}
  {\bibfnamefont {M.}~\bibnamefont {Sch{\"u}tz}},\ }\href {\doibase
  10.1103/PhysRevB.76.075101} {\bibfield  {journal} {\bibinfo  {journal} {Phys.
  Rev. B}\ }\textbf {\bibinfo {volume} {76}},\ \bibinfo {pages} {075101}
  (\bibinfo {year} {2007})}\BibitemShut {NoStop}%
\bibitem [{\citenamefont {Usvyat}\ \emph {et~al.}(2011)\citenamefont {Usvyat},
  \citenamefont {Civalleri}, \citenamefont {Maschio}, \citenamefont {Dovesi},
  \citenamefont {Pisani},\ and\ \citenamefont
  {Sch{\"u}tz}}]{usvyat_approaching_2011}%
  \BibitemOpen
  \bibfield  {author} {\bibinfo {author} {\bibfnamefont {D.}~\bibnamefont
  {Usvyat}}, \bibinfo {author} {\bibfnamefont {B.}~\bibnamefont {Civalleri}},
  \bibinfo {author} {\bibfnamefont {L.}~\bibnamefont {Maschio}}, \bibinfo
  {author} {\bibfnamefont {R.}~\bibnamefont {Dovesi}}, \bibinfo {author}
  {\bibfnamefont {C.}~\bibnamefont {Pisani}}, \ and\ \bibinfo {author}
  {\bibfnamefont {M.}~\bibnamefont {Sch{\"u}tz}},\ }\href {\doibase
  10.1063/1.3595514} {\bibfield  {journal} {\bibinfo  {journal} {J. Chem.
  Phys.}\ }\textbf {\bibinfo {volume} {134}},\ \bibinfo {pages} {214105}
  (\bibinfo {year} {2011})}\BibitemShut {NoStop}%
\bibitem [{\citenamefont {Hirata}\ \emph {et~al.}(2004)\citenamefont {Hirata},
  \citenamefont {Podeszwa}, \citenamefont {Tobita},\ and\ \citenamefont
  {Bartlett}}]{hirata_coupled-cluster_2004}%
  \BibitemOpen
  \bibfield  {author} {\bibinfo {author} {\bibfnamefont {S.}~\bibnamefont
  {Hirata}}, \bibinfo {author} {\bibfnamefont {R.}~\bibnamefont {Podeszwa}},
  \bibinfo {author} {\bibfnamefont {M.}~\bibnamefont {Tobita}}, \ and\ \bibinfo
  {author} {\bibfnamefont {R.~J.}\ \bibnamefont {Bartlett}},\ }\href {\doibase
  doi:10.1063/1.1637577} {\bibfield  {journal} {\bibinfo  {journal} {J. Chem.
  Phys.}\ }\textbf {\bibinfo {volume} {120}},\ \bibinfo {pages} {2581}
  (\bibinfo {year} {2004})}\BibitemShut {NoStop}%
\bibitem [{\citenamefont {Gr{\"u}neis}\ \emph {et~al.}(2010)\citenamefont
  {Gr{\"u}neis}, \citenamefont {Marsman},\ and\ \citenamefont
  {Kresse}}]{gruneis_second-order_2010}%
  \BibitemOpen
  \bibfield  {author} {\bibinfo {author} {\bibfnamefont {A.}~\bibnamefont
  {Gr{\"u}neis}}, \bibinfo {author} {\bibfnamefont {M.}~\bibnamefont
  {Marsman}}, \ and\ \bibinfo {author} {\bibfnamefont {G.}~\bibnamefont
  {Kresse}},\ }\href {\doibase 10.1063/1.3466765} {\bibfield  {journal}
  {\bibinfo  {journal} {J. Chem. Phys.}\ }\textbf {\bibinfo {volume} {133}},\
  \bibinfo {pages} {074107} (\bibinfo {year} {2010})}\BibitemShut {NoStop}%
\bibitem [{\citenamefont {Pino}\ and\ \citenamefont
  {Scuseria}(2004)}]{pino_importance_2004}%
  \BibitemOpen
  \bibfield  {author} {\bibinfo {author} {\bibfnamefont {R.}~\bibnamefont
  {Pino}}\ and\ \bibinfo {author} {\bibfnamefont {G.~E.}\ \bibnamefont
  {Scuseria}},\ }\href {\doibase 10.1063/1.1798991} {\bibfield  {journal}
  {\bibinfo  {journal} {J. Chem. Phys.}\ }\textbf {\bibinfo {volume} {121}},\
  \bibinfo {pages} {8113} (\bibinfo {year} {2004})}\BibitemShut {NoStop}%
\bibitem [{\citenamefont {Del~Ben}\ \emph {et~al.}(2012)\citenamefont
  {Del~Ben}, \citenamefont {Hutter},\ and\ \citenamefont
  {{VandeVondele}}}]{del_ben_second-order_2012}%
  \BibitemOpen
  \bibfield  {author} {\bibinfo {author} {\bibfnamefont {M.}~\bibnamefont
  {Del~Ben}}, \bibinfo {author} {\bibfnamefont {J.}~\bibnamefont {Hutter}}, \
  and\ \bibinfo {author} {\bibfnamefont {J.}~\bibnamefont {{VandeVondele}}},\
  }\href {\doibase 10.1021/ct300531w} {\bibfield  {journal} {\bibinfo
  {journal} {J. Chem. Theory Comput.}\ }\textbf {\bibinfo {volume} {8}},\
  \bibinfo {pages} {4177} (\bibinfo {year} {2012})}\BibitemShut {NoStop}%
\bibitem [{\citenamefont {Geckeis}\ \emph {et~al.}(2013)\citenamefont
  {Geckeis}, \citenamefont {L{\"u}tzenkirchen}, \citenamefont {Polly},
  \citenamefont {Rabung},\ and\ \citenamefont
  {Schmidt}}]{geckeis_mineralwater_2013}%
  \BibitemOpen
  \bibfield  {author} {\bibinfo {author} {\bibfnamefont {H.}~\bibnamefont
  {Geckeis}}, \bibinfo {author} {\bibfnamefont {J.}~\bibnamefont
  {L{\"u}tzenkirchen}}, \bibinfo {author} {\bibfnamefont {R.}~\bibnamefont
  {Polly}}, \bibinfo {author} {\bibfnamefont {T.}~\bibnamefont {Rabung}}, \
  and\ \bibinfo {author} {\bibfnamefont {M.}~\bibnamefont {Schmidt}},\ }\href
  {\doibase 10.1021/cr300370h} {\bibfield  {journal} {\bibinfo  {journal}
  {Chem. Rev.}\ }\textbf {\bibinfo {volume} {113}},\ \bibinfo {pages} {1016}
  (\bibinfo {year} {2013})}\BibitemShut {NoStop}%
\bibitem [{\citenamefont {Del~Ben}\ \emph {et~al.}(2013)\citenamefont
  {Del~Ben}, \citenamefont {Hutter},\ and\ \citenamefont
  {{VandeVondele}}}]{del_ben_electron_2013}%
  \BibitemOpen
  \bibfield  {author} {\bibinfo {author} {\bibfnamefont {M.}~\bibnamefont
  {Del~Ben}}, \bibinfo {author} {\bibfnamefont {J.}~\bibnamefont {Hutter}}, \
  and\ \bibinfo {author} {\bibfnamefont {J.}~\bibnamefont {{VandeVondele}}},\
  }\href {\doibase 10.1021/ct4002202} {\bibfield  {journal} {\bibinfo
  {journal} {J. Chem. Theory Comput.}\ }\textbf {\bibinfo {volume} {9}},\
  \bibinfo {pages} {2654} (\bibinfo {year} {2013})}\BibitemShut {NoStop}%
\bibitem [{\citenamefont {Usvyat}(2013)}]{usvyat_linear-scaling_2013}%
  \BibitemOpen
  \bibfield  {author} {\bibinfo {author} {\bibfnamefont {D.}~\bibnamefont
  {Usvyat}},\ }\href {\doibase 10.1063/1.4829898} {\bibfield  {journal}
  {\bibinfo  {journal} {J. Chem. Phys.}\ }\textbf {\bibinfo {volume} {139}},\
  \bibinfo {pages} {194101} (\bibinfo {year} {2013})}\BibitemShut {NoStop}%
\bibitem [{\citenamefont {M{\"u}ller}\ and\ \citenamefont
  {Usvyat}(2013)}]{muller_incrementally_2013}%
  \BibitemOpen
  \bibfield  {author} {\bibinfo {author} {\bibfnamefont {C.}~\bibnamefont
  {M{\"u}ller}}\ and\ \bibinfo {author} {\bibfnamefont {D.}~\bibnamefont
  {Usvyat}},\ }\href {\doibase 10.1021/ct400797w} {\bibfield  {journal}
  {\bibinfo  {journal} {J. Chem. Theory Comput.}\ }\textbf {\bibinfo {volume}
  {9}},\ \bibinfo {pages} {5590} (\bibinfo {year} {2013})}\BibitemShut
  {NoStop}%
\bibitem [{\citenamefont {M{\"u}ller}\ and\ \citenamefont
  {Paulus}(2012)}]{muller_wavefunction-based_2012}%
  \BibitemOpen
  \bibfield  {author} {\bibinfo {author} {\bibfnamefont {C.}~\bibnamefont
  {M{\"u}ller}}\ and\ \bibinfo {author} {\bibfnamefont {B.}~\bibnamefont
  {Paulus}},\ }\href {\doibase 10.1039/C2CP24020C} {\bibfield  {journal}
  {\bibinfo  {journal} {Phys. Chem. Chem. Phys.}\ }\textbf {\bibinfo {volume}
  {14}},\ \bibinfo {pages} {7605} (\bibinfo {year} {2012})}\BibitemShut
  {NoStop}%
\bibitem [{\citenamefont {Nolan}\ \emph {et~al.}(2010)\citenamefont {Nolan},
  \citenamefont {Bygrave}, \citenamefont {Allan},\ and\ \citenamefont
  {Manby}}]{nolan_comparison_2010}%
  \BibitemOpen
  \bibfield  {author} {\bibinfo {author} {\bibfnamefont {S.~J.}\ \bibnamefont
  {Nolan}}, \bibinfo {author} {\bibfnamefont {P.~J.}\ \bibnamefont {Bygrave}},
  \bibinfo {author} {\bibfnamefont {N.~L.}\ \bibnamefont {Allan}}, \ and\
  \bibinfo {author} {\bibfnamefont {F.~R.}\ \bibnamefont {Manby}},\ }\href
  {\doibase 10.1088/0953-8984/22/7/074201} {\bibfield  {journal} {\bibinfo
  {journal} {J. Phys.: Condens. Matter}\ }\textbf {\bibinfo {volume} {22}},\
  \bibinfo {pages} {074201} (\bibinfo {year} {2010})}\BibitemShut {NoStop}%
\bibitem [{\citenamefont {Gr{\"u}neis}\ \emph {et~al.}(2011)\citenamefont
  {Gr{\"u}neis}, \citenamefont {Booth}, \citenamefont {Marsman}, \citenamefont
  {Spencer}, \citenamefont {Alavi},\ and\ \citenamefont
  {Kresse}}]{gruneis_natural_2011}%
  \BibitemOpen
  \bibfield  {author} {\bibinfo {author} {\bibfnamefont {A.}~\bibnamefont
  {Gr{\"u}neis}}, \bibinfo {author} {\bibfnamefont {G.~H.}\ \bibnamefont
  {Booth}}, \bibinfo {author} {\bibfnamefont {M.}~\bibnamefont {Marsman}},
  \bibinfo {author} {\bibfnamefont {J.}~\bibnamefont {Spencer}}, \bibinfo
  {author} {\bibfnamefont {A.}~\bibnamefont {Alavi}}, \ and\ \bibinfo {author}
  {\bibfnamefont {G.}~\bibnamefont {Kresse}},\ }\href {\doibase
  10.1021/ct200263g} {\bibfield  {journal} {\bibinfo  {journal} {J. Chem.
  Theory Comput.}\ }\textbf {\bibinfo {volume} {7}},\ \bibinfo {pages} {2780}
  (\bibinfo {year} {2011})}\BibitemShut {NoStop}%
\bibitem [{\citenamefont {Booth}\ \emph {et~al.}(2013)\citenamefont {Booth},
  \citenamefont {Gr{\"u}neis}, \citenamefont {Kresse},\ and\ \citenamefont
  {Alavi}}]{booth_towards_2013}%
  \BibitemOpen
  \bibfield  {author} {\bibinfo {author} {\bibfnamefont {G.~H.}\ \bibnamefont
  {Booth}}, \bibinfo {author} {\bibfnamefont {A.}~\bibnamefont {Gr{\"u}neis}},
  \bibinfo {author} {\bibfnamefont {G.}~\bibnamefont {Kresse}}, \ and\ \bibinfo
  {author} {\bibfnamefont {A.}~\bibnamefont {Alavi}},\ }\href {\doibase
  10.1038/nature11770} {\bibfield  {journal} {\bibinfo  {journal} {Nature}\
  }\textbf {\bibinfo {volume} {493}},\ \bibinfo {pages} {365} (\bibinfo {year}
  {2013})}\BibitemShut {NoStop}%
\bibitem [{\citenamefont {Ke{\c c}eli}\ and\ \citenamefont
  {Hirata}(2010)}]{keceli_fast_2010}%
  \BibitemOpen
  \bibfield  {author} {\bibinfo {author} {\bibfnamefont {M.}~\bibnamefont
  {Ke{\c c}eli}}\ and\ \bibinfo {author} {\bibfnamefont {S.}~\bibnamefont
  {Hirata}},\ }\href {\doibase 10.1103/PhysRevB.82.115107} {\bibfield
  {journal} {\bibinfo  {journal} {Phys. Rev. B}\ }\textbf {\bibinfo {volume}
  {82}},\ \bibinfo {pages} {115107} (\bibinfo {year} {2010})}\BibitemShut
  {NoStop}%
\bibitem [{\citenamefont {Bartlett}\ and\ \citenamefont {Musia{\l
  }}(2007)}]{bartlett_coupled-cluster_2007}%
  \BibitemOpen
  \bibfield  {author} {\bibinfo {author} {\bibfnamefont {R.~J.}\ \bibnamefont
  {Bartlett}}\ and\ \bibinfo {author} {\bibfnamefont {M.}~\bibnamefont
  {Musia{\l }}},\ }\href {\doibase 10.1103/RevModPhys.79.291} {\bibfield
  {journal} {\bibinfo  {journal} {Rev. Mod. Phys.}\ }\textbf {\bibinfo {volume}
  {79}},\ \bibinfo {pages} {291} (\bibinfo {year} {2007})}\BibitemShut
  {NoStop}%
\bibitem [{\citenamefont {Bartlett}\ and\ \citenamefont
  {Silver}(1974)}]{bartlett_many-body_1974}%
  \BibitemOpen
  \bibfield  {author} {\bibinfo {author} {\bibfnamefont {R.~J.}\ \bibnamefont
  {Bartlett}}\ and\ \bibinfo {author} {\bibfnamefont {D.~M.}\ \bibnamefont
  {Silver}},\ }\href {\doibase 10.1016/0009-2614(74)85012-8} {\bibfield
  {journal} {\bibinfo  {journal} {Chem. Phys. Lett.}\ }\textbf {\bibinfo
  {volume} {29}},\ \bibinfo {pages} {199} (\bibinfo {year} {1974})}\BibitemShut
  {NoStop}%
\bibitem [{\citenamefont {Kelly}(1966)}]{kelly_many-body_1966}%
  \BibitemOpen
  \bibfield  {author} {\bibinfo {author} {\bibfnamefont {H.~P.}\ \bibnamefont
  {Kelly}},\ }\href {\doibase 10.1103/PhysRev.144.39} {\bibfield  {journal}
  {\bibinfo  {journal} {Phys. Rev.}\ }\textbf {\bibinfo {volume} {144}},\
  \bibinfo {pages} {39} (\bibinfo {year} {1966})}\BibitemShut {NoStop}%
\bibitem [{\citenamefont {Kelly}(1964)}]{kelly_many-body_1964}%
  \BibitemOpen
  \bibfield  {author} {\bibinfo {author} {\bibfnamefont {H.~P.}\ \bibnamefont
  {Kelly}},\ }\href {\doibase 10.1103/PhysRev.136.B896} {\bibfield  {journal}
  {\bibinfo  {journal} {Phys. Rev.}\ }\textbf {\bibinfo {volume} {136}},\
  \bibinfo {pages} {B896} (\bibinfo {year} {1964})}\BibitemShut {NoStop}%
\bibitem [{\citenamefont {Kelly}(1963)}]{kelly_correlation_1963}%
  \BibitemOpen
  \bibfield  {author} {\bibinfo {author} {\bibfnamefont {H.~P.}\ \bibnamefont
  {Kelly}},\ }\href {\doibase 10.1103/PhysRev.131.684} {\bibfield  {journal}
  {\bibinfo  {journal} {Phys. Rev.}\ }\textbf {\bibinfo {volume} {131}},\
  \bibinfo {pages} {684} (\bibinfo {year} {1963})}\BibitemShut {NoStop}%
\bibitem [{\citenamefont {Ostlund}\ and\ \citenamefont
  {Bowen}(1975)}]{ostlund_perturbation_1975}%
  \BibitemOpen
  \bibfield  {author} {\bibinfo {author} {\bibfnamefont {P.~N.~S.}\
  \bibnamefont {Ostlund}}\ and\ \bibinfo {author} {\bibfnamefont {M.~F.}\
  \bibnamefont {Bowen}},\ }\href {\doibase 10.1007/BF01135887} {\bibfield
  {journal} {\bibinfo  {journal} {Theoretica Chimica Acta}\ }\textbf {\bibinfo
  {volume} {40}},\ \bibinfo {pages} {175} (\bibinfo {year} {1975})}\BibitemShut
  {NoStop}%
\bibitem [{\citenamefont {Shepherd}\ \emph
  {et~al.}(2013{\natexlab{a}})\citenamefont {Shepherd}, \citenamefont
  {Henderson},\ and\ \citenamefont {Scuseria}}]{shepherd_range_2013}%
  \BibitemOpen
  \bibfield  {author} {\bibinfo {author} {\bibfnamefont {J.~J.}\ \bibnamefont
  {Shepherd}}, \bibinfo {author} {\bibfnamefont {T.~M.}\ \bibnamefont
  {Henderson}}, \ and\ \bibinfo {author} {\bibfnamefont {G.~E.}\ \bibnamefont
  {Scuseria}},\ }\href {http://arxiv.org/abs/1310.6425} {\emph {\bibinfo
  {title} {Range Separated Brueckner Coupled Cluster Doubles Theory}}},\
  \bibinfo {type} {{arXiv} e-print}\ \bibinfo {number} {1310.6425}\ (\bibinfo
  {year} {2013})\BibitemShut {NoStop}%
\bibitem [{\citenamefont {Roggero}\ \emph {et~al.}(2013)\citenamefont
  {Roggero}, \citenamefont {Mukherjee},\ and\ \citenamefont
  {Pederiva}}]{roggero_quantum_2013}%
  \BibitemOpen
  \bibfield  {author} {\bibinfo {author} {\bibfnamefont {A.}~\bibnamefont
  {Roggero}}, \bibinfo {author} {\bibfnamefont {A.}~\bibnamefont {Mukherjee}},
  \ and\ \bibinfo {author} {\bibfnamefont {F.}~\bibnamefont {Pederiva}},\
  }\href {\doibase 10.1103/PhysRevB.88.115138} {\bibfield  {journal} {\bibinfo
  {journal} {Phys. Rev. B}\ }\textbf {\bibinfo {volume} {88}},\ \bibinfo
  {pages} {115138} (\bibinfo {year} {2013})}\BibitemShut {NoStop}%
\bibitem [{\citenamefont {Bethe}(1956)}]{bethe_nuclear_1956}%
  \BibitemOpen
  \bibfield  {author} {\bibinfo {author} {\bibfnamefont {H.~A.}\ \bibnamefont
  {Bethe}},\ }\href {\doibase 10.1103/PhysRev.103.1353} {\bibfield  {journal}
  {\bibinfo  {journal} {Phys. Rev.}\ }\textbf {\bibinfo {volume} {103}},\
  \bibinfo {pages} {1353} (\bibinfo {year} {1956})}\BibitemShut {NoStop}%
\bibitem [{\citenamefont {Brueckner}\ \emph {et~al.}(1954)\citenamefont
  {Brueckner}, \citenamefont {Levinson},\ and\ \citenamefont
  {Mahmoud}}]{brueckner_two-body_1954}%
  \BibitemOpen
  \bibfield  {author} {\bibinfo {author} {\bibfnamefont {K.~A.}\ \bibnamefont
  {Brueckner}}, \bibinfo {author} {\bibfnamefont {C.~A.}\ \bibnamefont
  {Levinson}}, \ and\ \bibinfo {author} {\bibfnamefont {H.~M.}\ \bibnamefont
  {Mahmoud}},\ }\href {\doibase 10.1103/PhysRev.95.217} {\bibfield  {journal}
  {\bibinfo  {journal} {Phys. Rev.}\ }\textbf {\bibinfo {volume} {95}},\
  \bibinfo {pages} {217} (\bibinfo {year} {1954})}\BibitemShut {NoStop}%
\bibitem [{\citenamefont {Brueckner}(1954)}]{brueckner_nuclear_1954}%
  \BibitemOpen
  \bibfield  {author} {\bibinfo {author} {\bibfnamefont {K.~A.}\ \bibnamefont
  {Brueckner}},\ }\href {\doibase 10.1103/PhysRev.96.508} {\bibfield  {journal}
  {\bibinfo  {journal} {Phys. Rev.}\ }\textbf {\bibinfo {volume} {96}},\
  \bibinfo {pages} {508} (\bibinfo {year} {1954})}\BibitemShut {NoStop}%
\bibitem [{\citenamefont {Gr{\"u}neis}\ \emph {et~al.}(2009)\citenamefont
  {Gr{\"u}neis}, \citenamefont {Marsman}, \citenamefont {Harl}, \citenamefont
  {Schimka},\ and\ \citenamefont {Kresse}}]{gruneis_making_2009}%
  \BibitemOpen
  \bibfield  {author} {\bibinfo {author} {\bibfnamefont {A.}~\bibnamefont
  {Gr{\"u}neis}}, \bibinfo {author} {\bibfnamefont {M.}~\bibnamefont
  {Marsman}}, \bibinfo {author} {\bibfnamefont {J.}~\bibnamefont {Harl}},
  \bibinfo {author} {\bibfnamefont {L.}~\bibnamefont {Schimka}}, \ and\
  \bibinfo {author} {\bibfnamefont {G.}~\bibnamefont {Kresse}},\ }\href
  {\doibase 10.1063/1.3250347} {\bibfield  {journal} {\bibinfo  {journal} {J.
  Chem. Phys.}\ }\textbf {\bibinfo {volume} {131}},\ \bibinfo {pages} {154115}
  (\bibinfo {year} {2009})}\BibitemShut {NoStop}%
\bibitem [{\citenamefont {Harl}\ and\ \citenamefont
  {Kresse}(2008)}]{harl_cohesive_2008}%
  \BibitemOpen
  \bibfield  {author} {\bibinfo {author} {\bibfnamefont {J.}~\bibnamefont
  {Harl}}\ and\ \bibinfo {author} {\bibfnamefont {G.}~\bibnamefont {Kresse}},\
  }\href {\doibase 10.1103/PhysRevB.77.045136} {\bibfield  {journal} {\bibinfo
  {journal} {Phys. Rev. B}\ }\textbf {\bibinfo {volume} {77}},\ \bibinfo
  {pages} {045136} (\bibinfo {year} {2008})}\BibitemShut {NoStop}%
\bibitem [{\citenamefont {Harl}\ \emph {et~al.}(2010)\citenamefont {Harl},
  \citenamefont {Schimka},\ and\ \citenamefont {Kresse}}]{harl_assessing_2010}%
  \BibitemOpen
  \bibfield  {author} {\bibinfo {author} {\bibfnamefont {J.}~\bibnamefont
  {Harl}}, \bibinfo {author} {\bibfnamefont {L.}~\bibnamefont {Schimka}}, \
  and\ \bibinfo {author} {\bibfnamefont {G.}~\bibnamefont {Kresse}},\ }\href
  {\doibase 10.1103/PhysRevB.81.115126} {\bibfield  {journal} {\bibinfo
  {journal} {Phys. Rev. B}\ }\textbf {\bibinfo {volume} {81}},\ \bibinfo
  {pages} {115126} (\bibinfo {year} {2010})}\BibitemShut {NoStop}%
\bibitem [{\citenamefont {Scuseria}\ \emph {et~al.}(1986)\citenamefont
  {Scuseria}, \citenamefont {Lee},\ and\ \citenamefont
  {Schaefer~{III}}}]{scuseria_accelerating_1986}%
  \BibitemOpen
  \bibfield  {author} {\bibinfo {author} {\bibfnamefont {G.~E.}\ \bibnamefont
  {Scuseria}}, \bibinfo {author} {\bibfnamefont {T.~J.}\ \bibnamefont {Lee}}, \
  and\ \bibinfo {author} {\bibfnamefont {H.~F.}\ \bibnamefont
  {Schaefer~{III}}},\ }\href {\doibase 10.1016/0009-2614(86)80461-4} {\bibfield
   {journal} {\bibinfo  {journal} {Chem. Phys. Lett.}\ }\textbf {\bibinfo
  {volume} {130}},\ \bibinfo {pages} {236} (\bibinfo {year}
  {1986})}\BibitemShut {NoStop}%
\bibitem [{\citenamefont {Booth}\ \emph {et~al.}(2009)\citenamefont {Booth},
  \citenamefont {Thom},\ and\ \citenamefont {Alavi}}]{booth_fermion_2009}%
  \BibitemOpen
  \bibfield  {author} {\bibinfo {author} {\bibfnamefont {G.~H.}\ \bibnamefont
  {Booth}}, \bibinfo {author} {\bibfnamefont {A.~J.~W.}\ \bibnamefont {Thom}},
  \ and\ \bibinfo {author} {\bibfnamefont {A.}~\bibnamefont {Alavi}},\ }\href
  {\doibase 10.1063/1.3193710} {\bibfield  {journal} {\bibinfo  {journal} {J.
  Chem. Phys.}\ }\textbf {\bibinfo {volume} {131}},\ \bibinfo {pages} {054106}
  (\bibinfo {year} {2009})}\BibitemShut {NoStop}%
\bibitem [{\citenamefont {Cleland}\ \emph {et~al.}(2010)\citenamefont
  {Cleland}, \citenamefont {Booth},\ and\ \citenamefont
  {Alavi}}]{cleland_communications:_2010}%
  \BibitemOpen
  \bibfield  {author} {\bibinfo {author} {\bibfnamefont {D.}~\bibnamefont
  {Cleland}}, \bibinfo {author} {\bibfnamefont {G.~H.}\ \bibnamefont {Booth}},
  \ and\ \bibinfo {author} {\bibfnamefont {A.}~\bibnamefont {Alavi}},\ }\href
  {\doibase 10.1063/1.3302277} {\bibfield  {journal} {\bibinfo  {journal} {J.
  Chem. Phys.}\ }\textbf {\bibinfo {volume} {132}},\ \bibinfo {pages} {041103}
  (\bibinfo {year} {2010})}\BibitemShut {NoStop}%
\bibitem [{\citenamefont {Booth}\ \emph {et~al.}(2011)\citenamefont {Booth},
  \citenamefont {Cleland}, \citenamefont {Thom},\ and\ \citenamefont
  {Alavi}}]{booth_breaking_2011}%
  \BibitemOpen
  \bibfield  {author} {\bibinfo {author} {\bibfnamefont {G.~H.}\ \bibnamefont
  {Booth}}, \bibinfo {author} {\bibfnamefont {D.}~\bibnamefont {Cleland}},
  \bibinfo {author} {\bibfnamefont {A.~J.~W.}\ \bibnamefont {Thom}}, \ and\
  \bibinfo {author} {\bibfnamefont {A.}~\bibnamefont {Alavi}},\ }\href
  {\doibase 10.1063/1.3624383} {\bibfield  {journal} {\bibinfo  {journal} {J.
  Chem. Phys.}\ }\textbf {\bibinfo {volume} {135}},\ \bibinfo {pages} {084104}
  (\bibinfo {year} {2011})}\BibitemShut {NoStop}%
\bibitem [{\citenamefont {{L\'{o}pez R\'{\i}os}}(2013)}]{PabloPersComm}%
  \BibitemOpen
  \bibfield  {author} {\bibinfo {author} {\bibfnamefont {P.}~\bibnamefont
  {{L\'{o}pez R\'{\i}os}}},\ }\href@noop {} {}\bibinfo {type} {Personal
  communication}\ (\bibinfo {year} {2013})\BibitemShut {NoStop}%
\bibitem [{\citenamefont {Needs}\ \emph {et~al.}(2010)\citenamefont {Needs},
  \citenamefont {Towler}, \citenamefont {Drummond},\ and\ \citenamefont
  {L{\'o}pez~R{\'i}os}}]{needs_continuum_2010}%
  \BibitemOpen
  \bibfield  {author} {\bibinfo {author} {\bibfnamefont {R.~J.}\ \bibnamefont
  {Needs}}, \bibinfo {author} {\bibfnamefont {M.~D.}\ \bibnamefont {Towler}},
  \bibinfo {author} {\bibfnamefont {N.~D.}\ \bibnamefont {Drummond}}, \ and\
  \bibinfo {author} {\bibfnamefont {P.}~\bibnamefont {L{\'o}pez~R{\'i}os}},\
  }\href {\doibase 10.1088/0953-8984/22/2/023201} {\bibfield  {journal}
  {\bibinfo  {journal} {J. Phys.: Condens. Matter}\ }\textbf {\bibinfo {volume}
  {22}},\ \bibinfo {pages} {023201} (\bibinfo {year} {2010})}\BibitemShut
  {NoStop}%
\bibitem [{\citenamefont {Shepherd}\ \emph
  {et~al.}(2013{\natexlab{b}})\citenamefont {Shepherd}, \citenamefont
  {R{\'i}os}, \citenamefont {Drummond}, \citenamefont {Needs},\ and\
  \citenamefont {Alavi}}]{shepherd__2013}%
  \BibitemOpen
  \bibfield  {author} {\bibinfo {author} {\bibfnamefont {J.~J.}\ \bibnamefont
  {Shepherd}}, \bibinfo {author} {\bibfnamefont {P.~L.}\ \bibnamefont
  {R{\'i}os}}, \bibinfo {author} {\bibfnamefont {N.~D.}\ \bibnamefont
  {Drummond}}, \bibinfo {author} {\bibfnamefont {R.~J.}\ \bibnamefont {Needs}},
  \ and\ \bibinfo {author} {\bibfnamefont {A.}~\bibnamefont {Alavi}},\
  }\href@noop {} {}\bibinfo {type} {In preparation}\ (\bibinfo {year}
  {2013})\BibitemShut {NoStop}%
\bibitem [{\citenamefont {Foulkes}\ \emph {et~al.}(2001)\citenamefont
  {Foulkes}, \citenamefont {Mitas}, \citenamefont {Needs},\ and\ \citenamefont
  {Rajagopal}}]{foulkes_quantum_2001}%
  \BibitemOpen
  \bibfield  {author} {\bibinfo {author} {\bibfnamefont {W.~M.~C.}\
  \bibnamefont {Foulkes}}, \bibinfo {author} {\bibfnamefont {L.}~\bibnamefont
  {Mitas}}, \bibinfo {author} {\bibfnamefont {R.~J.}\ \bibnamefont {Needs}}, \
  and\ \bibinfo {author} {\bibfnamefont {G.}~\bibnamefont {Rajagopal}},\ }\href
  {\doibase 10.1103/RevModPhys.73.33} {\bibfield  {journal} {\bibinfo
  {journal} {Rev. Mod. Phys.}\ }\textbf {\bibinfo {volume} {73}},\ \bibinfo
  {pages} {33} (\bibinfo {year} {2001})}\BibitemShut {NoStop}%
\bibitem [{\citenamefont {Neese}\ \emph {et~al.}(2009)\citenamefont {Neese},
  \citenamefont {Wennmohs},\ and\ \citenamefont
  {Hansen}}]{neese_efficient_2009}%
  \BibitemOpen
  \bibfield  {author} {\bibinfo {author} {\bibfnamefont {F.}~\bibnamefont
  {Neese}}, \bibinfo {author} {\bibfnamefont {F.}~\bibnamefont {Wennmohs}}, \
  and\ \bibinfo {author} {\bibfnamefont {A.}~\bibnamefont {Hansen}},\ }\href
  {\doibase doi:10.1063/1.3086717} {\bibfield  {journal} {\bibinfo  {journal}
  {J. Chem. Phys.}\ }\textbf {\bibinfo {volume} {130}},\ \bibinfo {pages}
  {114108} (\bibinfo {year} {2009})}\BibitemShut {NoStop}%
\bibitem [{\citenamefont {Meyer}(1971)}]{meyer_ionization_1971}%
  \BibitemOpen
  \bibfield  {author} {\bibinfo {author} {\bibfnamefont {W.}~\bibnamefont
  {Meyer}},\ }\href {\doibase 10.1002/qua.560050839} {\bibfield  {journal}
  {\bibinfo  {journal} {Int. J. Quant. Chem.}\ }\textbf {\bibinfo {volume}
  {5}},\ \bibinfo {pages} {341{\textendash}348} (\bibinfo {year}
  {1971})}\BibitemShut {NoStop}%
\bibitem [{\citenamefont {Paldus}\ \emph {et~al.}(1984)\citenamefont {Paldus},
  \citenamefont {{\v C}{\'i}{\v z}ek},\ and\ \citenamefont
  {Takahashi}}]{paldus_approximate_1984}%
  \BibitemOpen
  \bibfield  {author} {\bibinfo {author} {\bibfnamefont {J.}~\bibnamefont
  {Paldus}}, \bibinfo {author} {\bibfnamefont {J.}~\bibnamefont {{\v C}{\'i}{\v
  z}ek}}, \ and\ \bibinfo {author} {\bibfnamefont {M.}~\bibnamefont
  {Takahashi}},\ }\href {\doibase 10.1103/PhysRevA.30.2193} {\bibfield
  {journal} {\bibinfo  {journal} {Phys. Rev. A}\ }\textbf {\bibinfo {volume}
  {30}},\ \bibinfo {pages} {2193} (\bibinfo {year} {1984})}\BibitemShut
  {NoStop}%
\bibitem [{\citenamefont {Bartlett}\ and\ \citenamefont {Musia{\l
  }}(2006)}]{bartlett_addition_2006}%
  \BibitemOpen
  \bibfield  {author} {\bibinfo {author} {\bibfnamefont {R.~J.}\ \bibnamefont
  {Bartlett}}\ and\ \bibinfo {author} {\bibfnamefont {M.}~\bibnamefont
  {Musia{\l }}},\ }\href {\doibase doi:10.1063/1.2387952} {\bibfield  {journal}
  {\bibinfo  {journal} {J. Chem. Phys.}\ }\textbf {\bibinfo {volume} {125}},\
  \bibinfo {pages} {204105} (\bibinfo {year} {2006})}\BibitemShut {NoStop}%
\bibitem [{\citenamefont {Huntington}\ and\ \citenamefont
  {Nooijen}(2010)}]{huntington_pccsd:_2010}%
  \BibitemOpen
  \bibfield  {author} {\bibinfo {author} {\bibfnamefont {L.~M.~J.}\
  \bibnamefont {Huntington}}\ and\ \bibinfo {author} {\bibfnamefont
  {M.}~\bibnamefont {Nooijen}},\ }\href {\doibase doi:10.1063/1.3494113}
  {\bibfield  {journal} {\bibinfo  {journal} {J. Chem. Phys.}\ }\textbf
  {\bibinfo {volume} {133}},\ \bibinfo {pages} {184109} (\bibinfo {year}
  {2010})}\BibitemShut {NoStop}%
\bibitem [{\citenamefont {Masur}\ \emph {et~al.}(2013)\citenamefont {Masur},
  \citenamefont {Usvyat},\ and\ \citenamefont
  {Sch{\"u}tz}}]{masur_efficient_2013}%
  \BibitemOpen
  \bibfield  {author} {\bibinfo {author} {\bibfnamefont {O.}~\bibnamefont
  {Masur}}, \bibinfo {author} {\bibfnamefont {D.}~\bibnamefont {Usvyat}}, \
  and\ \bibinfo {author} {\bibfnamefont {M.}~\bibnamefont {Sch{\"u}tz}},\
  }\href {\doibase 10.1063/1.4826534} {\bibfield  {journal} {\bibinfo
  {journal} {J. Chem. Phys.}\ }\textbf {\bibinfo {volume} {139}},\ \bibinfo
  {pages} {164116} (\bibinfo {year} {2013})}\BibitemShut {NoStop}%
\bibitem [{\citenamefont {Kats}\ and\ \citenamefont
  {Manby}(2013)}]{kats_communication:_2013}%
  \BibitemOpen
  \bibfield  {author} {\bibinfo {author} {\bibfnamefont {D.}~\bibnamefont
  {Kats}}\ and\ \bibinfo {author} {\bibfnamefont {F.~R.}\ \bibnamefont
  {Manby}},\ }\href {\doibase doi:10.1063/1.4813481} {\bibfield  {journal}
  {\bibinfo  {journal} {J. Chem. Phys.}\ }\textbf {\bibinfo {volume} {139}},\
  \bibinfo {pages} {021102} (\bibinfo {year} {2013})}\BibitemShut {NoStop}%
\bibitem [{\citenamefont {Scuseria}(1995)}]{scuseria_connections_1995}%
  \BibitemOpen
  \bibfield  {author} {\bibinfo {author} {\bibfnamefont {G.~E.}\ \bibnamefont
  {Scuseria}},\ }\href {\doibase 10.1002/qua.560550211} {\bibfield  {journal}
  {\bibinfo  {journal} {Int. J. Quant. Chem.}\ }\textbf {\bibinfo {volume}
  {55}},\ \bibinfo {pages} {165{\textendash}171} (\bibinfo {year}
  {1995})}\BibitemShut {NoStop}%
\bibitem [{\citenamefont {Bartlett}(2009)}]{bartlett_towards_2009}%
  \BibitemOpen
  \bibfield  {author} {\bibinfo {author} {\bibfnamefont {R.~J.}\ \bibnamefont
  {Bartlett}},\ }\href {\doibase 10.1016/j.cplett.2009.10.053} {\bibfield
  {journal} {\bibinfo  {journal} {Chem. Phys. Lett.}\ }\textbf {\bibinfo
  {volume} {484}},\ \bibinfo {pages} {1} (\bibinfo {year} {2009})}\BibitemShut
  {NoStop}%
\bibitem [{\citenamefont {Watts}\ and\ \citenamefont
  {Bartlett}(1994)}]{watts_coupled-cluster_1994}%
  \BibitemOpen
  \bibfield  {author} {\bibinfo {author} {\bibfnamefont {J.~D.}\ \bibnamefont
  {Watts}}\ and\ \bibinfo {author} {\bibfnamefont {R.~J.}\ \bibnamefont
  {Bartlett}},\ }\href {\doibase 10.1002/qua.560520820} {\bibfield  {journal}
  {\bibinfo  {journal} {Int. J. Quant. Chem.}\ }\textbf {\bibinfo {volume}
  {52}},\ \bibinfo {pages} {195{\textendash}203} (\bibinfo {year}
  {1994})}\BibitemShut {NoStop}%
\bibitem [{\citenamefont {Sherrill}\ \emph {et~al.}(1998)\citenamefont
  {Sherrill}, \citenamefont {Krylov}, \citenamefont {Byrd},\ and\ \citenamefont
  {Head-Gordon}}]{sherrill_energies_1998}%
  \BibitemOpen
  \bibfield  {author} {\bibinfo {author} {\bibfnamefont {C.~D.}\ \bibnamefont
  {Sherrill}}, \bibinfo {author} {\bibfnamefont {A.~I.}\ \bibnamefont
  {Krylov}}, \bibinfo {author} {\bibfnamefont {E.~F.~C.}\ \bibnamefont {Byrd}},
  \ and\ \bibinfo {author} {\bibfnamefont {M.}~\bibnamefont {Head-Gordon}},\
  }\href {\doibase doi:10.1063/1.477023} {\bibfield  {journal} {\bibinfo
  {journal} {J. Chem. Phys.}\ }\textbf {\bibinfo {volume} {109}},\ \bibinfo
  {pages} {4171} (\bibinfo {year} {1998})}\BibitemShut {NoStop}%
\bibitem [{\citenamefont {Hampel}\ \emph {et~al.}(1992)\citenamefont {Hampel},
  \citenamefont {Peterson},\ and\ \citenamefont
  {Werner}}]{hampel_comparison_1992}%
  \BibitemOpen
  \bibfield  {author} {\bibinfo {author} {\bibfnamefont {C.}~\bibnamefont
  {Hampel}}, \bibinfo {author} {\bibfnamefont {K.~A.}\ \bibnamefont
  {Peterson}}, \ and\ \bibinfo {author} {\bibfnamefont {H.-J.}\ \bibnamefont
  {Werner}},\ }\href {\doibase 10.1016/0009-2614(92)86093-W} {\bibfield
  {journal} {\bibinfo  {journal} {Chem. Phys. Lett.}\ }\textbf {\bibinfo
  {volume} {190}},\ \bibinfo {pages} {1} (\bibinfo {year} {1992})}\BibitemShut
  {NoStop}%
\bibitem [{\citenamefont {Scuseria}(1994)}]{scuseria_alternative_1994}%
  \BibitemOpen
  \bibfield  {author} {\bibinfo {author} {\bibfnamefont {G.~E.}\ \bibnamefont
  {Scuseria}},\ }\href {\doibase 10.1016/0009-2614(94)00747-0} {\bibfield
  {journal} {\bibinfo  {journal} {Chem. Phys. Lett.}\ }\textbf {\bibinfo
  {volume} {226}},\ \bibinfo {pages} {251} (\bibinfo {year}
  {1994})}\BibitemShut {NoStop}%
\bibitem [{\citenamefont {Moussa}(2014)}]{moussa_cubic-scaling_2014}%
  \BibitemOpen
  \bibfield  {author} {\bibinfo {author} {\bibfnamefont {J.~E.}\ \bibnamefont
  {Moussa}},\ }\href {\doibase 10.1063/1.4855255} {\bibfield  {journal}
  {\bibinfo  {journal} {J. Chem. Phys.}\ }\textbf {\bibinfo {volume} {140}},\
  \bibinfo {pages} {014107} (\bibinfo {year} {2014})}\BibitemShut {NoStop}%
\bibitem [{\citenamefont {Baguet}\ \emph {et~al.}(2013)\citenamefont {Baguet},
  \citenamefont {Delyon}, \citenamefont {Bernu},\ and\ \citenamefont
  {Holzmann}}]{baguet_hartree-fock_2013}%
  \BibitemOpen
  \bibfield  {author} {\bibinfo {author} {\bibfnamefont {L.}~\bibnamefont
  {Baguet}}, \bibinfo {author} {\bibfnamefont {F.}~\bibnamefont {Delyon}},
  \bibinfo {author} {\bibfnamefont {B.}~\bibnamefont {Bernu}}, \ and\ \bibinfo
  {author} {\bibfnamefont {M.}~\bibnamefont {Holzmann}},\ }\href {\doibase
  10.1103/PhysRevLett.111.166402} {\bibfield  {journal} {\bibinfo  {journal}
  {Phys. Rev. Lett.}\ }\textbf {\bibinfo {volume} {111}},\ \bibinfo {pages}
  {166402} (\bibinfo {year} {2013})}\BibitemShut {NoStop}%
\bibitem [{\citenamefont {Zhang}\ and\ \citenamefont
  {Ceperley}(2008)}]{zhang_hartree-fock_2008}%
  \BibitemOpen
  \bibfield  {author} {\bibinfo {author} {\bibfnamefont {S.}~\bibnamefont
  {Zhang}}\ and\ \bibinfo {author} {\bibfnamefont {D.~M.}\ \bibnamefont
  {Ceperley}},\ }\href {\doibase 10.1103/PhysRevLett.100.236404} {\bibfield
  {journal} {\bibinfo  {journal} {Phys. Rev. Lett.}\ }\textbf {\bibinfo
  {volume} {100}},\ \bibinfo {pages} {236404} (\bibinfo {year}
  {2008})}\BibitemShut {NoStop}%
\bibitem [{\citenamefont {Drummond}\ \emph {et~al.}(2004)\citenamefont
  {Drummond}, \citenamefont {Radnai}, \citenamefont {Trail}, \citenamefont
  {Towler},\ and\ \citenamefont {Needs}}]{drummond_diffusion_2004}%
  \BibitemOpen
  \bibfield  {author} {\bibinfo {author} {\bibfnamefont {N.~D.}\ \bibnamefont
  {Drummond}}, \bibinfo {author} {\bibfnamefont {Z.}~\bibnamefont {Radnai}},
  \bibinfo {author} {\bibfnamefont {J.~R.}\ \bibnamefont {Trail}}, \bibinfo
  {author} {\bibfnamefont {M.~D.}\ \bibnamefont {Towler}}, \ and\ \bibinfo
  {author} {\bibfnamefont {R.~J.}\ \bibnamefont {Needs}},\ }\href {\doibase
  10.1103/PhysRevB.69.085116} {\bibfield  {journal} {\bibinfo  {journal} {Phys.
  Rev. B}\ }\textbf {\bibinfo {volume} {69}},\ \bibinfo {pages} {085116}
  (\bibinfo {year} {2004})}\BibitemShut {NoStop}%
\bibitem [{\citenamefont {Onsager}\ \emph {et~al.}(1966)\citenamefont
  {Onsager}, \citenamefont {Mittag},\ and\ \citenamefont
  {Stephen}}]{onsager_integrals_1966}%
  \BibitemOpen
  \bibfield  {author} {\bibinfo {author} {\bibfnamefont {L.}~\bibnamefont
  {Onsager}}, \bibinfo {author} {\bibfnamefont {L.}~\bibnamefont {Mittag}}, \
  and\ \bibinfo {author} {\bibfnamefont {M.~J.}\ \bibnamefont {Stephen}},\
  }\href {\doibase 10.1002/andp.19664730108} {\bibfield  {journal} {\bibinfo
  {journal} {Annalen der Physik}\ }\textbf {\bibinfo {volume} {473}},\ \bibinfo
  {pages} {71{\textendash}77} (\bibinfo {year} {1966})}\BibitemShut {NoStop}%
\bibitem [{\citenamefont {Freeman}(1983)}]{freeman_coupled-cluster_1983}%
  \BibitemOpen
  \bibfield  {author} {\bibinfo {author} {\bibfnamefont {D.~L.}\ \bibnamefont
  {Freeman}},\ }\href {\doibase 10.1088/0022-3719/16/4/017} {\bibfield
  {journal} {\bibinfo  {journal} {Journal of Physics C: Solid State Physics}\
  }\textbf {\bibinfo {volume} {16}},\ \bibinfo {pages} {711} (\bibinfo {year}
  {1983})}\BibitemShut {NoStop}%
\bibitem [{\citenamefont {Cioslowski}\ and\ \citenamefont
  {Ziesche}(2005)}]{cioslowski_applicability_2005}%
  \BibitemOpen
  \bibfield  {author} {\bibinfo {author} {\bibfnamefont {J.}~\bibnamefont
  {Cioslowski}}\ and\ \bibinfo {author} {\bibfnamefont {P.}~\bibnamefont
  {Ziesche}},\ }\href {\doibase 10.1103/PhysRevB.71.125105} {\bibfield
  {journal} {\bibinfo  {journal} {Phys. Rev. B}\ }\textbf {\bibinfo {volume}
  {71}},\ \bibinfo {pages} {125105} (\bibinfo {year} {2005})}\BibitemShut
  {NoStop}%
\bibitem [{\citenamefont {Yurtsever}\ and\ \citenamefont
  {Tanatar}(2001)}]{yurtsever_ladder_2001}%
  \BibitemOpen
  \bibfield  {author} {\bibinfo {author} {\bibfnamefont {A.}~\bibnamefont
  {Yurtsever}}\ and\ \bibinfo {author} {\bibfnamefont {B.}~\bibnamefont
  {Tanatar}},\ }\href {\doibase 10.1103/PhysRevB.63.125320} {\bibfield
  {journal} {\bibinfo  {journal} {Phys. Rev. B}\ }\textbf {\bibinfo {volume}
  {63}},\ \bibinfo {pages} {125320} (\bibinfo {year} {2001})}\BibitemShut
  {NoStop}%
\bibitem [{\citenamefont {Kecke}\ and\ \citenamefont
  {H{\"a}usler}(2004)}]{kecke_ladder_2004}%
  \BibitemOpen
  \bibfield  {author} {\bibinfo {author} {\bibfnamefont {L.}~\bibnamefont
  {Kecke}}\ and\ \bibinfo {author} {\bibfnamefont {W.}~\bibnamefont
  {H{\"a}usler}},\ }\href {\doibase 10.1103/PhysRevB.69.085103} {\bibfield
  {journal} {\bibinfo  {journal} {Phys. Rev. B}\ }\textbf {\bibinfo {volume}
  {69}},\ \bibinfo {pages} {085103} (\bibinfo {year} {2004})}\BibitemShut
  {NoStop}%
\bibitem [{\citenamefont {Hirata}(2011)}]{hirata_thermodynamic_2011}%
  \BibitemOpen
  \bibfield  {author} {\bibinfo {author} {\bibfnamefont {S.}~\bibnamefont
  {Hirata}},\ }\href {\doibase 10.1007/s00214-011-0954-4} {\bibfield  {journal}
  {\bibinfo  {journal} {Theor. Chem. Acc.}\ }\textbf {\bibinfo {volume}
  {129}},\ \bibinfo {pages} {727} (\bibinfo {year} {2011})}\BibitemShut
  {NoStop}%
\bibitem [{\citenamefont {Shepherd}\ and\ \citenamefont
  {Gr{\"u}neis}(2012)}]{shepherd_correlation_2012}%
  \BibitemOpen
  \bibfield  {author} {\bibinfo {author} {\bibfnamefont {J.~J.}\ \bibnamefont
  {Shepherd}}\ and\ \bibinfo {author} {\bibfnamefont {A.}~\bibnamefont
  {Gr{\"u}neis}},\ }\href {http://arxiv.org/abs/1208.6103} {\emph {\bibinfo
  {title} {Correlation Energy Divergences in Metallic Systems}}},\ \bibinfo
  {type} {{arXiv} e-print}\ \bibinfo {number} {1208.6103}\ (\bibinfo {year}
  {2012})\BibitemShut {NoStop}%
\bibitem [{Note1()}]{Note1}%
  \BibitemOpen
  \bibinfo {note} {For precise values of $M$ the reader is referred to
  Ref.~\protect \rev@citealp {shepherd_correlation_2012}}\BibitemShut {NoStop}%
\bibitem [{\citenamefont {Campillo}\ \emph {et~al.}(1999)\citenamefont
  {Campillo}, \citenamefont {Pitarke}, \citenamefont {Rubio}, \citenamefont
  {Zarate},\ and\ \citenamefont {Echenique}}]{campillo_inelastic_1999}%
  \BibitemOpen
  \bibfield  {author} {\bibinfo {author} {\bibfnamefont {I.}~\bibnamefont
  {Campillo}}, \bibinfo {author} {\bibfnamefont {J.~M.}\ \bibnamefont
  {Pitarke}}, \bibinfo {author} {\bibfnamefont {A.}~\bibnamefont {Rubio}},
  \bibinfo {author} {\bibfnamefont {E.}~\bibnamefont {Zarate}}, \ and\ \bibinfo
  {author} {\bibfnamefont {P.~M.}\ \bibnamefont {Echenique}},\ }\href {\doibase
  10.1103/PhysRevLett.83.2230} {\bibfield  {journal} {\bibinfo  {journal}
  {Phys. Rev. Lett.}\ }\textbf {\bibinfo {volume} {83}},\ \bibinfo {pages}
  {2230} (\bibinfo {year} {1999})}\BibitemShut {NoStop}%
\bibitem [{\citenamefont {Kurth}\ and\ \citenamefont
  {Perdew}(1999)}]{kurth_density-functional_1999}%
  \BibitemOpen
  \bibfield  {author} {\bibinfo {author} {\bibfnamefont {S.}~\bibnamefont
  {Kurth}}\ and\ \bibinfo {author} {\bibfnamefont {J.~P.}\ \bibnamefont
  {Perdew}},\ }\href {\doibase 10.1103/PhysRevB.59.10461} {\bibfield  {journal}
  {\bibinfo  {journal} {Phys. Rev. B}\ }\textbf {\bibinfo {volume} {59}},\
  \bibinfo {pages} {10461} (\bibinfo {year} {1999})}\BibitemShut {NoStop}%
\bibitem [{\citenamefont {Oshikiri}\ and\ \citenamefont
  {Aryasetiawan}(1999)}]{oshikiri_band_1999}%
  \BibitemOpen
  \bibfield  {author} {\bibinfo {author} {\bibfnamefont {M.}~\bibnamefont
  {Oshikiri}}\ and\ \bibinfo {author} {\bibfnamefont {F.}~\bibnamefont
  {Aryasetiawan}},\ }\href {\doibase 10.1103/PhysRevB.60.10754} {\bibfield
  {journal} {\bibinfo  {journal} {Phys. Rev. B}\ }\textbf {\bibinfo {volume}
  {60}},\ \bibinfo {pages} {10754} (\bibinfo {year} {1999})}\BibitemShut
  {NoStop}%
\bibitem [{\citenamefont {Nesbet}(1958)}]{nesbet_brueckners_1958}%
  \BibitemOpen
  \bibfield  {author} {\bibinfo {author} {\bibfnamefont {R.~K.}\ \bibnamefont
  {Nesbet}},\ }\href {\doibase 10.1103/PhysRev.109.1632} {\bibfield  {journal}
  {\bibinfo  {journal} {Phys. Rev.}\ }\textbf {\bibinfo {volume} {109}},\
  \bibinfo {pages} {1632} (\bibinfo {year} {1958})}\BibitemShut {NoStop}%
\bibitem [{Note2()}]{Note2}%
  \BibitemOpen
  \bibinfo {note} {This has been investigated explicitly by looking at the
  exact wavefunction in Ref.~\protect \rev@citealp
  {shepherd_quantum_2013}}\BibitemShut {NoStop}%
\bibitem [{\citenamefont {Shepherd}(2013)}]{shepherd_quantum_2013}%
  \BibitemOpen
  \bibfield  {author} {\bibinfo {author} {\bibfnamefont {J.~J.}\ \bibnamefont
  {Shepherd}},\ }\emph {\bibinfo {title} {A Quantum Chemical Perspective on the
  Homogeneous Electron Gas}},\ \href@noop {} {Ph.D. thesis},\ \bibinfo
  {school} {University of Cambridge} (\bibinfo {year} {2013})\BibitemShut
  {NoStop}%
\end{thebibliography}
%

\end{document}